\documentclass{article}

\pdfoutput=1
\usepackage{arxiv}

\usepackage[utf8]{inputenc} 
\usepackage[T1]{fontenc}    
\usepackage{hyperref}       
\usepackage{url}            
\usepackage{booktabs}       
\usepackage{amsfonts}       
\usepackage{nicefrac}       
\usepackage{microtype}      
\usepackage{enumitem}
\usepackage{graphicx}
\usepackage{titlesec}
\graphicspath{ {./Figures/} }
\usepackage{natbib}
\usepackage{doi}
\usepackage[symbol]{footmisc}
\usepackage{caption}
\usepackage{amssymb} 

\titleformat{\paragraph}
{\normalfont\normalsize\bfseries}{\theparagraph}{1em}{}
\titlespacing*{\paragraph}
{0pt}{3.25ex plus 1ex minus .2ex}{1.5ex plus .2ex}

\title{How habitable are M-dwarf Exoplanets? Modeling surface conditions and exploring the role of melanins in the survival of \emph{Aspergillus niger} spores under exoplanet-like radiation}

\renewcommand{\shorttitle}{\textit{Assessing Habitability on M-Dwarf Exoplanets with Aspergillus niger Spores}}

\hypersetup{
pdftitle={MotaEtAl2024},
pdfkeywords={Exoplanets, M-dwarfs, Radiation, Aspergillus niger, Habitability, Melanin},
}

\begin{document}

\begin{center}
\rule{\linewidth}{0.5mm} \\[0.4cm]
{{\linespread{1.3}\Large\mdseries\MakeUppercase{How habitable are M-dwarf Exoplanets? Modeling surface conditions and exploring the role of melanins in the survival of \emph{Aspergillus niger} spores under exoplanet-like radiation}\par} 
\rule{\linewidth}{0.5mm} \\[0.5cm]

    \texttt{SUBMITTED TO \textit{ASTROBIOLOGY}} \\
    \vspace{0.1cm}
    \texttt{6\textsuperscript{TH} OF MARCH 2024}

\vspace{0.25cm}

\textbf{Afonso Mota\textsuperscript{1,2}\footnote[1]{Corresponding author. Email: \href{mailto:afonsomota.papers@gmail.com}{\texttt{afonsomota.papers@gmail.com}}}, Stella Koch\textsuperscript{1}, Daniel Matthiae\textsuperscript{3}, Nuno Santos\textsuperscript{2,4}, and Marta Cortesão\textsuperscript{1}} \\ \vspace{0.3cm}
\textit{\textsuperscript{1}Aerospace Microbiology Research Group, Institute of Aerospace Medicine, German Aerospace Center (DLR), Cologne, Germany \\ \vspace{0.2cm}
\textsuperscript{2}Instituto de Astrofísica e Ciências do Espaço, Universidade do Porto, CAUP, Rua das Estrelas, 4150-762 Porto, Portugal \\ \vspace{0.2cm}
\textsuperscript{3}Biophysics Research Group, Institute of Aerospace Medicine, German Aerospace Center (DLR), Cologne, Germany \\ \vspace{0.2cm}
\textsuperscript{4}Departamento de Física e Astronomia, Faculdade de Ciências, Universidade do Porto, Rua do Campo Alegre, 4169-007 Porto, Portugal \\}}
\end{center}
\vspace{0.5cm}

\begin{abstract}
    Exoplanet habitability remains a challenging field due to the large distances separating Earth from other stars. Using insights from biology and astrophysics, we studied the habitability of M-dwarf exoplanets by modeling their surface temperature and flare UV and X-ray doses using the Martian atmosphere as a shielding model. Analyzing the Proxima Centauri and TRAPPIST-1 systems, our models suggest that Proxima b and TRAPPIST-1 e are likeliest to have temperatures compatible with surface liquid water, as well as tolerable radiation environments.
    Results of the modeling were used as a basis for microbiology experiments to assess spore survival of the melanin-rich fungus \emph{Aspergillus niger} to exoplanet-like radiation (UV-C and X-rays). Results showed that \textit{A. niger} spores can endure superflare events on M-dwarf planets when shielded by a Mars-like atmosphere or by a thin layer of soil or water. Melanin-deficient spores suspended in a melanin-rich solution showed higher survival rates and germination efficiency when compared to melanin-free solutions. Overall, the models developed in this work establish a framework for microbiological research in habitability studies. Finally, we showed that \textit{A. niger} spores can survive harsh radiation conditions of simulated exoplanets, also emphasizing the importance of multifunctional molecules like melanins in radiation shielding beyond Earth. 

\end{abstract}


\section{Introduction}
    Conventionally, exoplanet habitability is assessed solely with its orbital distance to its star, and whether that orbit is present within the star’s expected “Goldilocks”, or “habitable”, zone \citep{kasting1993}. However, this is a reductionistic description of habitability – not only does it depend on many other factors, but the definition focuses only on Earth-like planets, leaving out cases where planets and moons outside this zone may harbor habitable conditions, both in our Solar System \citep{nimmo2016} and beyond \citep{madhusudhan2021, seager2013}.
    Exoplanets are notoriously difficult to observe and study due to the large distances that separate the Earth from even the closest stars. That said, clear developments have been made towards the identification of rocky planets orbiting M-dwarf stars, the smallest, coolest, and most common stars in the Universe. Current methods employed for exoplanet detection predominantly revolve around indirect strategies like radial velocity and transit methods \citep{santos2020}. Future studies will be able to characterize exoplanet atmospheres, and this is considered one of the main avenues for the possible detection of biosignatures \citep{palle2023}. Moreover, telescopes are a constantly improving technology, and techniques used to analyze observational data are also in continuous evolution. In the next few years, many more planets will be revealed in the Milky Way, some of which could have habitable conditions where we may be able to find indications of the presence of life \citep{rauer2014}. 
    
    The study of exoplanet habitability is a critical aspect of astrobiology, as it provides insights into the potential existence of life outside of Earth, within and beyond our solar system. Both the planet’s surface temperature and radiation environment play a crucial role in determining habitability. Ultraviolet (UV) radiation and X-rays can be detrimental to potential organisms on the surface, as well as alter a planet’s atmospheric composition, particularly for planets orbiting M-dwarf stars, which have increased activity and strong flares \citep{howard2018, tilley2019, yamashiki2019}.
    
    Microorganisms are key objects of study in astrobiology since these are the most widespread, abundant, and adaptable life forms on our planet. Microbes span across the three domains of life: Bacteria, Archaea, and Eukarya. Their diversity is not only taxonomic but also manifests in an array of diverse metabolic capabilities and lifestyles, which allows them to inhabit a considerable variety of environments \citep{brock2003}. The study of microbial extremophiles is particularly relevant in astrobiology, as it can provide information about the potential for life in extraterrestrial environments that may resemble these extreme conditions on Earth (e.g. \citealp{cortesao2020, pacelli2020}).
    
    Within the microbial world, pigments are a diverse group of compounds that contribute to the survival of many extremophiles, exhibiting a remarkable range of colors, structures, and functions. These serve various roles in cellular processes such as thermoregulation, quenching oxidative stress, and cellular messaging \citep{malik2012}. Pigments have been proposed as promising biosignatures for exoplanets \citep{coelho2022, schwieterman2015}. Melanins, in particular, are found in both eukaryotic and prokaryotic microorganisms alike, enabling some organisms to survive in extreme environments \citep{cordero2017}. It has been proposed that melanins or similar pigments could have been key for the origin and development of life on Earth \citep{dischia2021}, and perhaps on other worlds. 
    
    In the context of astrobiology, and particularly astromycology, the study of extremotolerant fungi has proven critical to better understanding the limits of life and habitability. The field has advanced significantly through experiments such as LIFE \citep{onofri2015} and the subsequent BIOMEX experiment \citep{devera2019}. \textit{Aspergillus niger}, an extremotolerant filamentous fungus, has been frequently used as a model organism for studying fungal survival in extreme environments, growing in a wide range of conditions \citep{cortesao2020}. \textit{A. niger} spores have a dense and complex melanin coating which increases their resistance to many stresses, such as UV and X-ray radiation, and oxidative stress \citep{cortesao2020, xu2022}. \textit{A. niger} has also been found to be present in space stations, highlighting its endurance to the conditions of space \citep{cortesao2021, romsdahl2018}. Furthermore, \textit{A. niger} has been extensively studied as a model organism for biotechnology and microbiology \citep{cairns2018}.
    
    Terrestrial microorganisms can inform the potential for the habitability of exoplanets. This study presents an interdisciplinary approach, bridging astrophysics and microbiology, to approach the problem of what might constitute a habitable M-dwarf exoplanet. We present a model for estimating the surface temperature of an exoplanet, as well as its radiation environment assuming Mars-like atmospheric properties, due to the red planet’s high astrobiological relevance but limited current habitability. On exoplanets estimated to have dayside temperatures amenable to life, we assess their potential survivability by exposing the model fungus \textit{A. niger} to the modeled radiation conditions (UV and X-rays) and examine the role of solubilized melanin in enhancing spore resistance and germination.

\section{Materials and Methods}
\label{sec:headings}

\subsection{Selection of Exoplanet Systems and Planetary Parameters}

    To provide a testbed for exoplanet habitability, this work focused on two astrobiology relevant star systems – Proxima Centauri (referred to hereafter simply as Proxima) and TRAPPIST-1. The studied rocky exoplanets were Proxima b and d, and TRAPPIST-1 d, e and f, due to their potential habitability, and, in the case of Proxima b and d, vicinity to Earth and recent discovery. Proxima b is thought to be a somewhat cool Earth-like planet with a tidally locked or 3:2 resonant orbit \citep{anglada-escude2016, boutle2017, sergeev2020, suarez-mascareno2020, turbet2016}. Proxima d is a small sub-Earth discovered in 2022, which is likely dry and barren \citep{faria2022}. TRAPPIST-1 d, e, and f are inside their star’s optimistic “Goldilocks” zone. TRAPPIST-1 e is thought to be one of the exoplanets found so far that is most likely to have conditions to be habitable \citep{krissansen-totton2022,quarles2017,sergeev2020}. Finally, TRAPPIST-1 f has an Earth-like mass and radius \citep{agol2021}. Although previous research suggested it may have limited habitability due to a massive water-rich envelope with surface temperature and pressure too high for liquid water \citep{quarles2017}, recent findings have indicated that it may have a cold but habitable climate, perhaps with liquid water on its surface \citep{krissansen-totton2022}. Table 1 shows the relevant planetary parameters used in this study, with the Solar System’s rocky planets included as examples. Stellar temperatures \citep{agol2021, pavlenko2017, williamssun2022} and planet semi-major axes \citep{agol2021, faria2022, williamsplanetary2023} were taken from the literature and used to calculate the incident flux for each exoplanet.

\begin{table}
	\centering
	\includegraphics{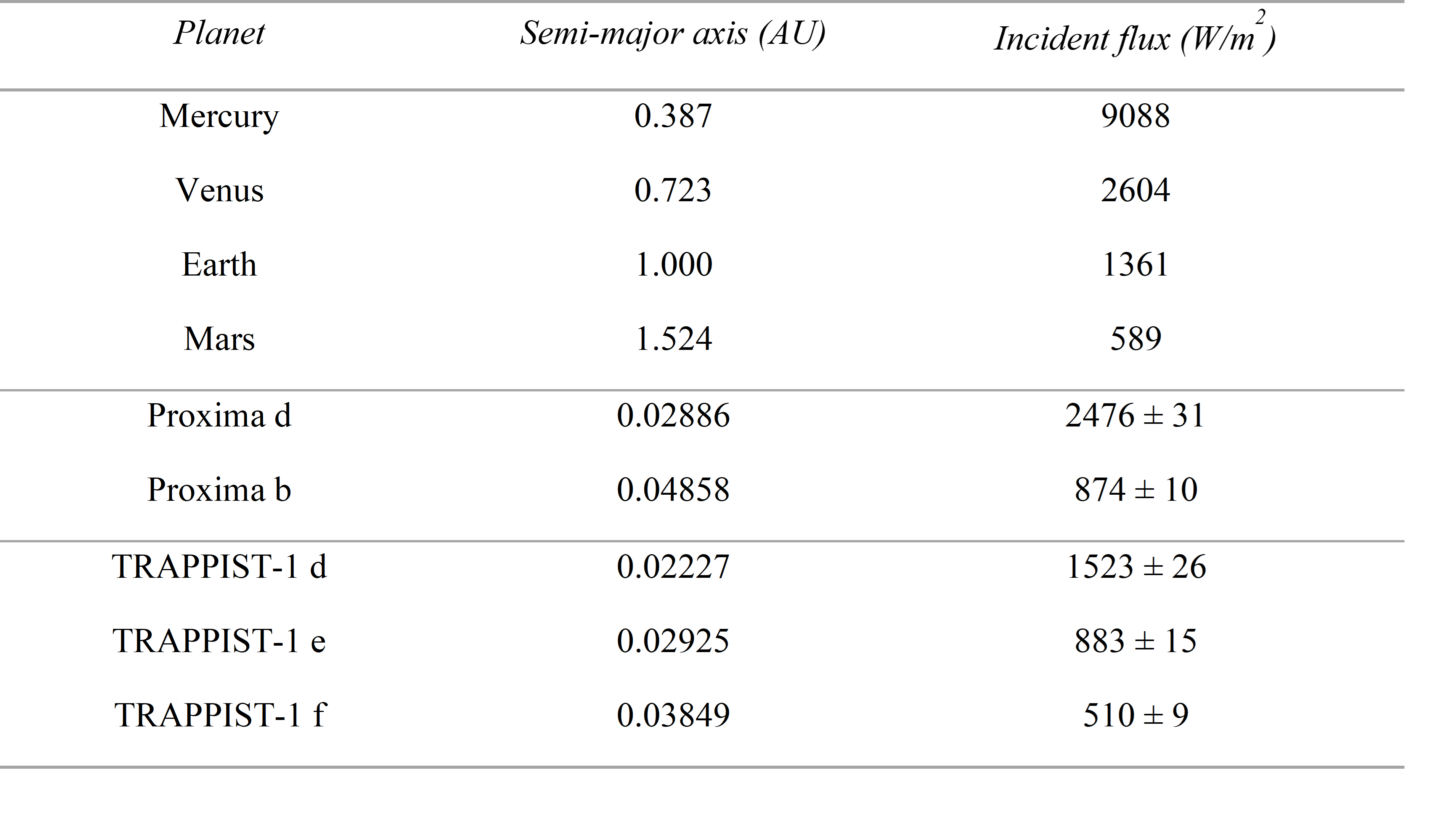}
	\caption{Incident flux calculated as $F=\frac{\sigma R_{\star}^2T_{\star}^4}{a^2}$, where $T_{\star}$ and $R_{\star}$ are the star’s temperature and radius \citep{agol2021, pavlenko2017, williamssun2022}, respectively, and $a$ is the planet’s semi-major axis \citep{agol2021, faria2022, williamsplanetary2023}. $\sigma$ is the Stefan-Boltzmann constant.}
	\label{tab:tab1}
\end{table}

\subsection{Modelling of Exoplanet Conditions}
\subsubsection{Estimating Rocky Planet Surface Temperatures}

    A crucial factor controlling habitability is the planet’s surface temperature, which is difficult to determine since it is influenced by many factors, such as albedo and greenhouse effect \citep{seager2011, sergeev2020}. However, for exoplanets, most of these factors are not easily known. Therefore, a more commonly used model is equilibrium temperature ($T_{eq}$), a theoretical temperature value following several assumptions like the absence of an atmosphere and a perfect heat distribution (\citep{kump2014}. Naturally, this is not necessarily a good representation of the surface temperature of the planet, and complex models are frequently employed to study possible scenarios and the corresponding surface temperature distributions (e.g. \citealp{boutle2017}). However, a simple model remains to be developed that considers factors such as atmospheric greenhouse effect, energy distribution efficiency, and planetary reflectivity (albedo), without requiring the detailed knowledge that most climate models need (\citealp{godolt2016, lincowski2018}), but still producing accurate general predictions when compared to the equilibrium temperature.
    
    We developed a simple equation based on the $T_{eq}$ which can be employed to better estimate the surface temperature of rocky exoplanets depending on several factors. First, we considered a factor $\varepsilon$ representing the atmospheric greenhouse effect, variable between 0 (no greenhouse, e.g. Mercury) and 1 (strong greenhouse, e.g. Venus); as well as $N$, the number of simulated atmospheric layers. Earth has a $\varepsilon$ value of 0.77-0.79 \citep{jaboc1999, liu2020}. For very powerful greenhouses, like Venus-like planets, $N >>$ 1, while for all other cases, $N =$ 1 \citep{liu2020}. To account for orbital resonance patterns like tidal locking, we added a variable $f$ (0.5 $< f <$ 1) representing the energy distribution efficiency of a planet, as proposed in previous research \citep{seager2011}. In tidally locked planets with no energy redistribution, $f =$ 0.5. For planets with efficient heat diffusion, $f =$ 1. A final model equation for the dayside temperature can be written as:

\begin{equation}
	T_d=\sqrt[4]{\frac {N \cdot F(1-A_B)}{4f\sigma(1-\frac{\varepsilon}{2})}}=T_{eq} \cdot \sqrt[4]{\frac {N}{f(1-\frac{\varepsilon}{2})}}
\end{equation}

    $F$, the energy flux that reaches the planet; $\sigma$, the Stefan-Boltzmann constant; and $A_B$, the Bond albedo of the planet; are used to calculate the $T_{eq}$. A more detailed description of the derivation process can be found in the Supplementary Material.

\subsubsection{X-ray Fluxes at the Top-of-atmosphere of M-dwarf Exoplanets}


    A stellar flare is a sudden and dramatic increase in a star's brightness caused by a large-scale magnetic event, which releases a substantial amount of energy across the electromagnetic spectrum, including X-rays and UV light. The average expected X-ray energy released during a standard M-dwarf flare is around 2.5$\times$10$^{30}$ erg \citep{welsh2007}. We considered this value when simulating an average flare from Proxima; however, since TRAPPIST-1 produces generally less energetic flares \citep{maas2022, yamashiki2019}, we assume a mean of 1.0$\times$10$^{30}$ erg of X-rays released by a conventional flare. Hence, the total estimated energy per unit area that reaches a certain planet’s top-of-atmosphere (TOA) can be calculated using the inverse-square law applied to its orbital semi-major axis ($a$):

\begin{equation}
    E_p(\frac{J}{m^2})=E_{flare}(erg) \times 10^{-7} \times \frac{1}{4 \pi a^2}
\end{equation}

   Here, the 10$^{-7}$ factor serves to convert the flare energy from erg to Joule, as flare energies are traditionally presented in erg. Although spectra of the X-ray photon energy distributions of M-dwarf stellar flares have not yet been acquired (to the best of our knowledge), it is expected that these will follow a power-law, similar to the Sun’s X-ray emission during a flare. The Sun’s X-ray photon output (1–100 keV) during observed flares has been compiled in previous studies, such as in \citet{maggio2008} and \citet{inglis2009}. Regression analysis of both sets of data was performed and used to fit power-law equations to model the photon release during flares (Fig. 1).


\begin{figure}
	\centering
	\includegraphics[scale=0.8]{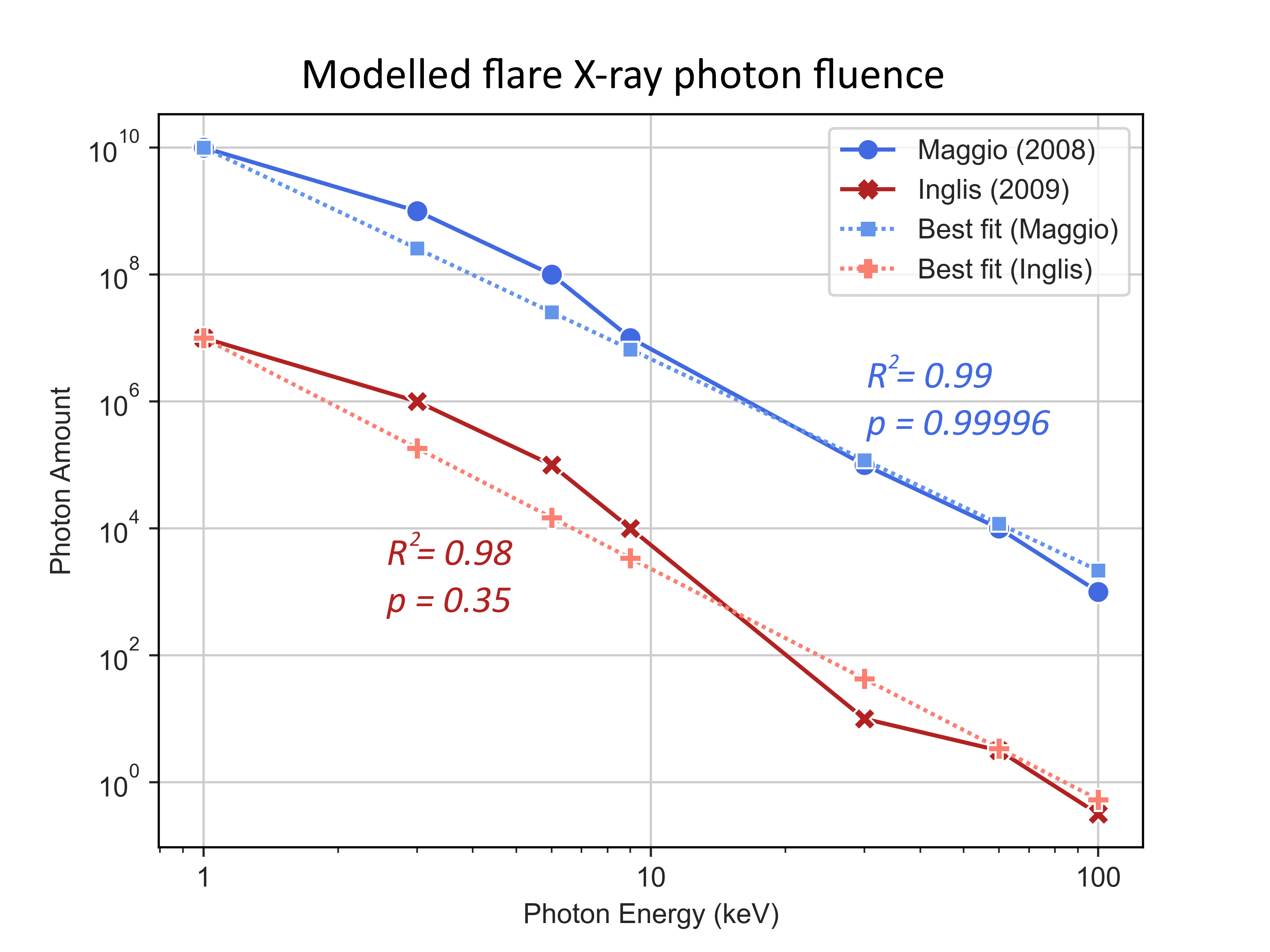}
	\caption{Photon fluence (photons cm$^{-2}$ keV$^{-1}$) in the X-ray band during an average   solar flare as shown in \citet{maggio2008}, over 100 seconds, and \citet{inglis2009} over 1 second. \textit{p}-values shown are from Kolmogorov-Smirnov tests and indicate that the fit equations are adequate representations of the data.}
	\label{fig:fig1}
\end{figure}
    

    To create a general model for an average flare, we took the mean of the exponents from both equations, getting $k=$ –3.4845. Hence, the best-fit equation to calculate the stellar photon fluence in the X-ray wavelength range (1–100 keV) is:

\begin{equation}
    F(E)=F_1 \times E^{-3.4845}
\end{equation}

    Here, $E$ is the photon energy (in keV), and $F_1$ represents the initial condition of the power-law, that is, the absolute flux of photons with $E=$ 1 keV. Therefore, $F_1$ is variable depending on the considered flare energy, and its value is larger for stronger flares. The total X-ray energy received by a planet during a flare, $E_p$, can be calculated by considering the stellar X-ray photon fluence $F(E)$, determined in equation (3), as an integration within the X-ray energy wavelengths ($E$, from 1–100 keV):

\begin{equation}
    E_p=\int_{1}^{100}E \times F(E) \,dE
\end{equation}

    When integrated numerically and rearranged, this equation gives the X-ray photon fluence on the planet, $\theta(E)$, in photons m$^{-2}$ keV$^{-1}$, as:  

\begin{equation}
    \theta(E)=1.4022 \times E_p \times E^{-3.4845}
\end{equation}

    The $\theta(E)$ represents the number of photons for each energy value, and therefore is useful to calculate the dose throughout the X-ray wavelengths. X-rays are measured in absorbed dose (Gy, J/kg), and the same X-ray flux can lead to different absorbed doses based on the irradiated material. Conversion from energy flux to absorbed dose is not direct, but an estimation can be done using the target’s mass-energy absorption coefficient ($\frac {\mu_{en}}{\rho}$) \citep{hubbell1995}. The most accurate materials to use as a reference to microbes are liquid water and soft tissue (ICRU-44), with similar $\frac {\mu_{en}}{\rho}$.

    Assuming an approximately constant fluence over the target volume, the absorbed dose ($D(E)$) for photons of energy E can be calculated with the previously described $\frac {\mu_{en}}{\rho}$ values, without attenuation, using the following factor $\delta$:

\begin{equation}
    \delta(E)=E \cdot \frac {\mu_{en}}{\rho}(E) \cdot \theta(E)
\end{equation}

    However, since $\frac {\mu_{en}}{\rho}$ data comes from discrete data points, we must consider the distance between tabled energy values ($\Delta E_i$) in the calculations, as well as the average $\delta$ between successive data points. Therefore, the equation for determining $D(E)$ emerges as:

\begin{equation}
    D(E)=(E_{i+1}-E_i) \cdot \frac{\delta(E_i)+\delta(E_{i+1})}{2}
\end{equation}

    Equation (7) yields the dose values in keV/kg. This result can then be converted to J/kg.
    
\paragraph{Atmospheric Attenuation of X-rays and (Sub-)Surface Doses}

    Additionally, to model the physical attenuation of X-rays passing through a planet’s atmosphere, surface, or water layers, we used Mars as a model planet for its astrobiological relevance, taking compositions of Martian soil simulant (MGS-1) from literature \citep{cannon2019}, and assuming an atmospheric content 95\% CO\textsubscript{2} and 5\% N\textsubscript{2}. The interaction of X-rays with the Martian atmosphere has been previously modeled \citep{smith2007}, from which we calculated the fraction of transmitted photons over several wavelengths by dividing the surface photon amount by the incident photon amount at the TOA. These values are plotted in Fig. 2. Then, through least squares fitting, we obtain the equation for the atmospheric transmission fraction ($T_A$) over the X-ray wavelengths (1-100 keV), which follows the function $T_A(E)=\frac{0.2579}{1+e^{-0.1102 \cdot (E-41.54)}) }-0.0194$ (Fig. 2). For surface attenuation (water and Mars-like soil), mass attenuation ($\frac{\mu}{\rho}$; \citealp{berger1998}) values were used. X-ray transmittance through a material follows the law $T_S(E)=e^{-\frac{\mu}{\rho}(E) \cdot x \cdot \rho}$, where $x$ is the depth traversed by the photons, and $\rho$ the material density – 1000 kg/m$^3$ for water and 1290 kg/m$^3$ for Mars soil \citep{cannon2019} – resulting in an estimation of the surface attenuation ($T_S$).

Factoring in both $T_A$ and $T_S$ yields a final model for estimating the surface or sub-surface X-ray absorbed dose of a Mars-like model exoplanet:

\begin{figure}
	\centering
	\includegraphics[scale=0.8]{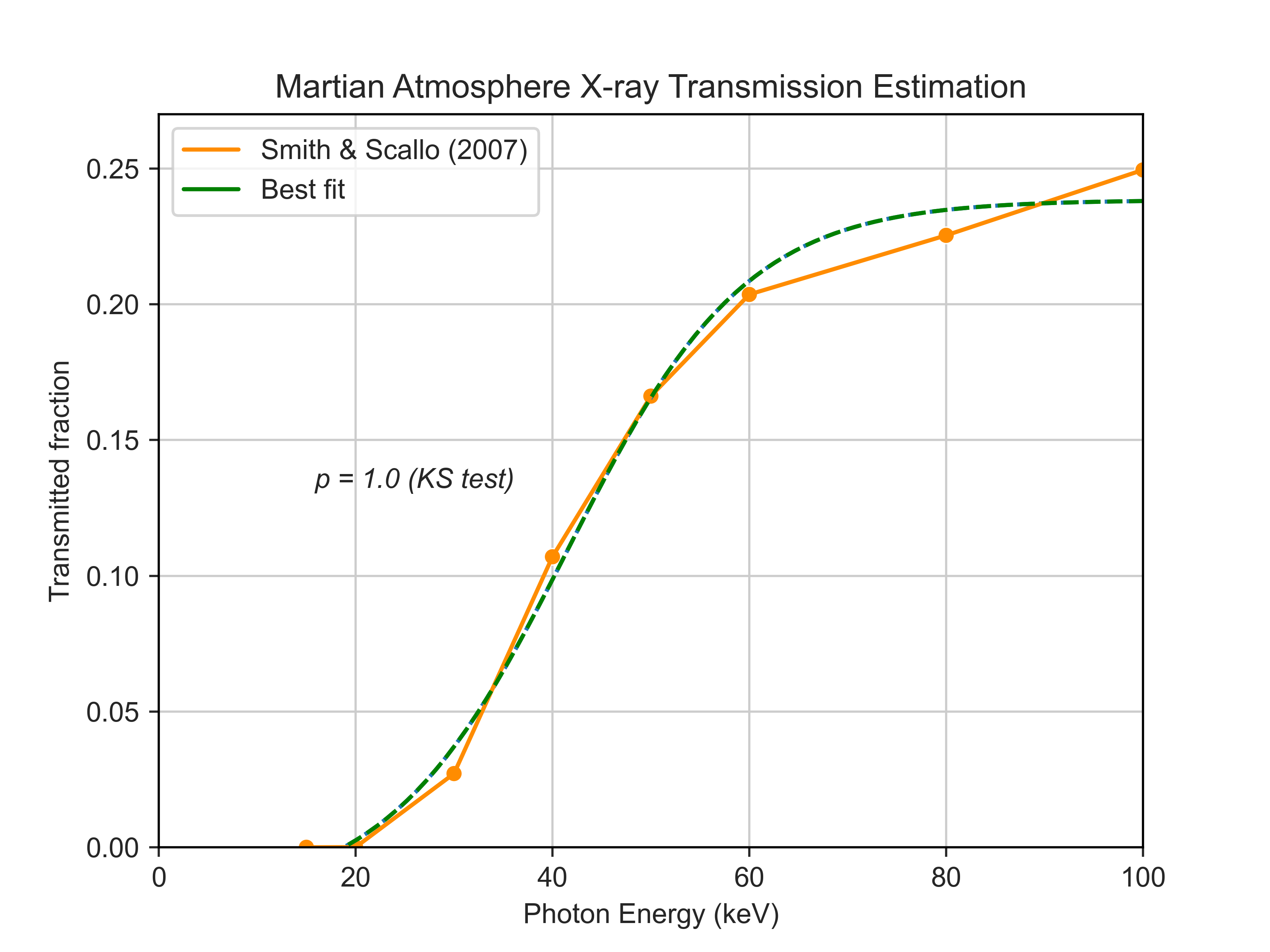}
	\caption{Modeled X-ray transmittance of the Martian atmosphere based on data from \citet{smith2007}. The data points taken were those for a stronger flare (spectral index \textit{p} = 2.5) at an atmospheric column density of 16 g/cm$^2$. As shown in the figure, the Kolmogorov-Smirnov test (\textit{p} = 1.0) indicates that the fitted curve is an adequate model of the data. }
	\label{fig:fig2}
\end{figure}

\begin{equation}
    \hat{D}(E)=D(E) \cdot T_A(E) \cdot T_S(E)
\end{equation}

\subsubsection{UV Fluxes at the Top-of-atmosphere and Surface of M-Dwarf Exoplanets}

    M-dwarf flare energies in the UV range are similar to the X-ray fluxes \citep{welsh2007}. However, unlike for X-rays, the UV flux of M-dwarfs during flares tends to be similar across the UV wavelength range, with only a slight increase from lower to higher wavelengths \citep{ranjan2017, segura2010, tilley2019}. Therefore, for this work, we assumed a constant flux across the UV range for flares of Proxima Centauri and TRAPPIST-1. Furthermore, the UV spectrum is divided into UV-A (315-400 nm), UV-B (280-315 nm), and UV-C (100-280 nm), where UV-C is the most harmful \citep{bucheli-witschel2010}. Since the UV flux is uniform across wavelengths, the estimated fractions of each UV type are 28.3\% UV-A, 11.7\% UV-B, and 60.0\% UV-C. 
    We modeled the UV attenuation by the Martian atmosphere using data from \citet{cockell2000b} (Fig. 3). UV light below 200 nm is completely attenuated by the Martian atmosphere.

\begin{figure}
	\centering
	\includegraphics[scale=0.8]{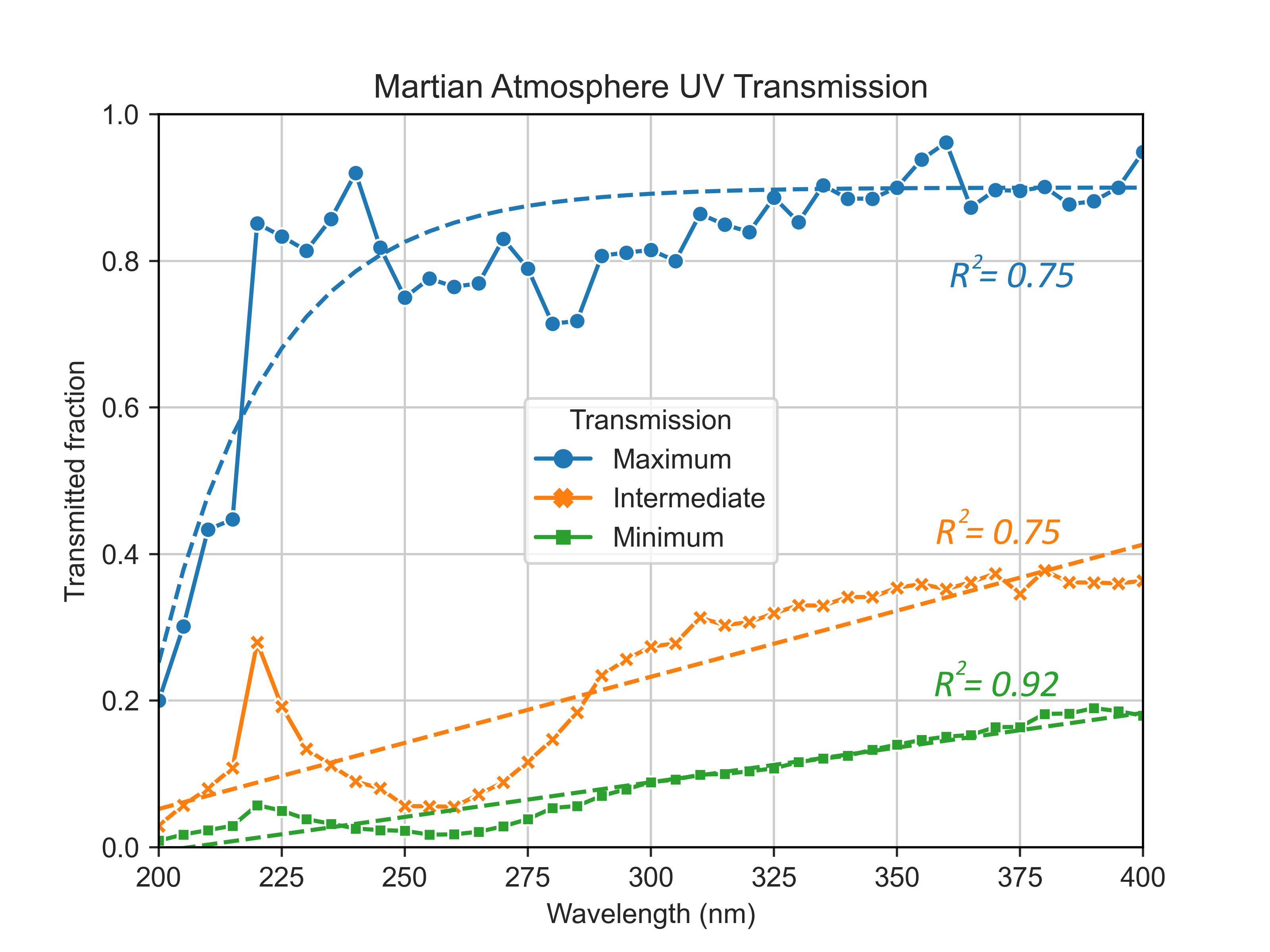}
	\caption{Dots and straight lines represent the Martian atmosphere’s UV transmission fraction calculated from \citet{cockell2000b}. Maximum transmission conditions involve clear skies (little suspended dust) and vertical photon flux (shorter atmospheric interaction distance); intermediate conditions are clear skies with light rays hitting the planet at a 60º angle; minimum transmission comprises a dusty atmosphere and 60º ray angle. Dashed lines are calculated curves fitting the data with corresponding Pearson’s correlation values ($R^2$).}
	\label{fig:fig3}
\end{figure}

\subsection{Estimating the Survival of Model Organisms to the Calculated Radiation Doses}

    After calculating the surface UV and X-ray dose for each planet using the modeled transmittance curves, we compared these results to the LD$_{90}$ (D$_{10}$) values – the radiation dose required to kill 90\% of a population of a certain microorganism – of three model microbes. The selected organisms were \textit{Escherichia coli} (mesophile), \textit{Aspergillus niger} (extremotolerant, high UV resistance), and \textit{Deinococcus radiodurans} (polyextremophile, high X-ray resistance). \textit{E. coli} LD$_{90}$’s are 22.6 J/m$^2$ for UV \citep{gascon1995} and 200 Gy for X-rays \citep{moreira2012}. \textit{A. niger} LD$_{90}$'s are 1038 J/m$^2$ for UV and 366 Gy for X-rays \citep{cortesao2020}. \textit{D. radiodurans} LD$_{90}$’s are 533 J/m$^2$ for UV \citep{gascon1995} and 1.6$\times$10$^4$ Gy for X-rays \citep{slade2011}.

\subsection{Experimental Setup: Organisms, Media, Pigments}

    After creating models for studying rocky exoplanets, several microbiology experiments were conducted to test how fungal spores may endure exoplanet-like radiation. Three strains of \textit{\textit{A. niger}} were used in this work: a wild-type strain, N402; a DHN-melanin deficient mutant strain, MA93.1 \citep{cortesao2020, jorgensen2011}; and a mutant strain modified to produce and excrete pyomelanin, OS4.3 \citep{koch2023}. Spores were collected from 3-day-old cultures grown in complete medium (CM), prepared as shown in \citet{cortesao2020}. at 30°C. After this, radiation exposure and viability assays were carried out using a minimal medium (MM) as described in \citet{koch2023}. All experiments were conducted using of three technical replicates per strain (n = 3).

\subsubsection{Pyomelanin Solution Preparation}

    The melanin production process was adapted from \citet{koch2023}, after which the pyomelanin-rich supernatant was filtered through a sterile Miracloth filter to eliminate any remaining hyphal fragments. This filtered solution was then stored at 4ºC until it was used to suspend the spores for irradiation experiments. As controls, we used the standard 0.9\% NaCl saline solution, as well as a N402-derived supernatant produced similarly to the process described above.

\subsubsection{Radiation Exposure Conditions}

    For UV and X-ray exposure, exoplanet-like doses were obtained by employing the models developed during this study, as described in sections 2.2.2 and 2.2.3. Experiments were conducted assuming two types of flare events for the considered M-dwarfs, standard flares and superflares. The used UV doses were 1000, 2500, and 5000 J/m$^2$, in addition to non-irradiated controls. \textit{\textit{A. niger}} spore irradiation was done as described in \citet{cortesao2020}.

    For UV, spores were subjected to UV radiation in Petri dishes, with each sample containing 15 mL of a suspension of 10$^6$ spores/mL in 0.9\% NaCl saline solution. At this concentration, spores form a monolayer, and there is no additional protection caused by a high cell density ($>$10$^7$ spores/mL for larger volumes). A UV lamp (VL-215-LC, Vilbert Lourmat, SN.: 14 100595) with a monochromatic UV-C wavelength of 254 nm was used for the irradiation process. Magnetic stirrers were utilized to continuously mix the spore suspension during exposure, preventing the settling of the spores on the bottom and the resulting mutual shielding among them. Spore viability and growth capability were assayed as shown in 2.3.3.
    
    For X-rays, \textit{A. niger} spores were suspended in PCR tubes (Brand) containing 100 $\mu$L of 0.9\% NaCl saline solution at a concentration of 10$^7$ spores/mL. A higher concentration was used when compared to UV irradiation due to the reduced volume of the spore suspension, ensuring the presence of a monolayer, and thus no spore-to-spore protection, even at a higher cell density. The RS225 X-ray device (Gulmay Medical Systems, Camberley, Surrey, UK) was used for irradiation, operating unfiltered at 200 kV and 15 mA, enabling high-dose exposure in a short period. The X-ray machine outputs photons on a spectrum up to $>$100 keV, with the largest peak at around 60 keV. The dose rate (in Gy/min) was determined using the UNIDOS webline and a TM30013 ionization chamber (PTW, Freiburg, Germany), allowing the calculation of the correct exposure time to achieve the desired X-ray absorbed doses for the samples. The used X-ray doses were 100, 500, and 1000 Gy, in addition to non-irradiated controls.
    
\subsubsection{Survival and Growth Assays}

    The survival and viability of \textit{A. niger} spores were assessed by testing their ability to form colonies following exposure to the experimental conditions. Serial dilutions of irradiated spore samples were prepared up to 10$^{-6}$ in a 96-well plate, with a total volume of 100 $\mu$L per well. To determine the number of colony-forming units (CFUs), 20 $\mu$L of each dilution was plated in on 1/6 of a Petri dish containing MM agar supplemented with 0.05\% Triton X-100 to reduce colony size and aid in counting. After a 2-day incubation at 30°C, colonies were counted, and the survival fraction ratio ($\frac{N}{N_0}$) was calculated. Here, $N$ represents the CFU for treated samples, while $N_0$ is the CFU for the controls. 

    Additionally, spore survival and growth profile were further assessed via live-cell imaging using the oCelloScope\texttrademark (BioSense Solutions ApS, Farum, Denmark, \citealp{koch2023}). To prepare the samples for observation of germination and hyphal formation, spore samples were diluted to a concentration of 10$^5$ spores/mL and incubated in liquid MM at 22°C over a 48-56 hour period. The oCelloScope\texttrademark analyzed the changing fungal biomass over time for each well using the built-in SESAfungi algorithm normalized at 4 hours after inoculation in the medium, determined as the necessary amount of time for settling of spores (without germination), dust, and other suspended particles.

    To compare the growth profile results generated by the oCelloScope\texttrademark, the Mann-Whitney (M-W), and Kruskal-Wallis (K-W) tests were performed, due to the non-parametric nature of the data. For survival fraction evaluations (parametric data), t-tests were used. The assumed significance threshold value was \textit{p} = 0.05.

\section{Results}
\subsection{Dayside Surface Temperatures}

    Running the model from equation (1) for a selected planet yields a dayside surface temperature matrix, with the calculated temperature depending on the greenhouse effect strength, Bond albedo, and energy distribution efficiency. Temperature matrices for Proxima d, Proxima b, TRAPPIST-1 e and TRAPPIST-1 f are shown in Fig. 4. The colormap was normalized with a minimum value of 235 K (-38 ºC) and a maximum value of 340 K (67 ºC) since modeling shows that surface temperatures below 235 K or above 340 K cannot sustain liquid water under any circumstances \citep{godolt2016}. A surface temperature between 273 K and 313 K (0 to 40 ºC) has been suggested as being ideal to maximize habitability \citep{godolt2016}. Results for TRAPPIST-1 d are also shown in the Supplementary Material (Fig. S2) for a thick Venus-like atmosphere and for a Mars or Earth-like atmosphere.

\begin{figure}
	\centering
	\includegraphics[scale=0.6]{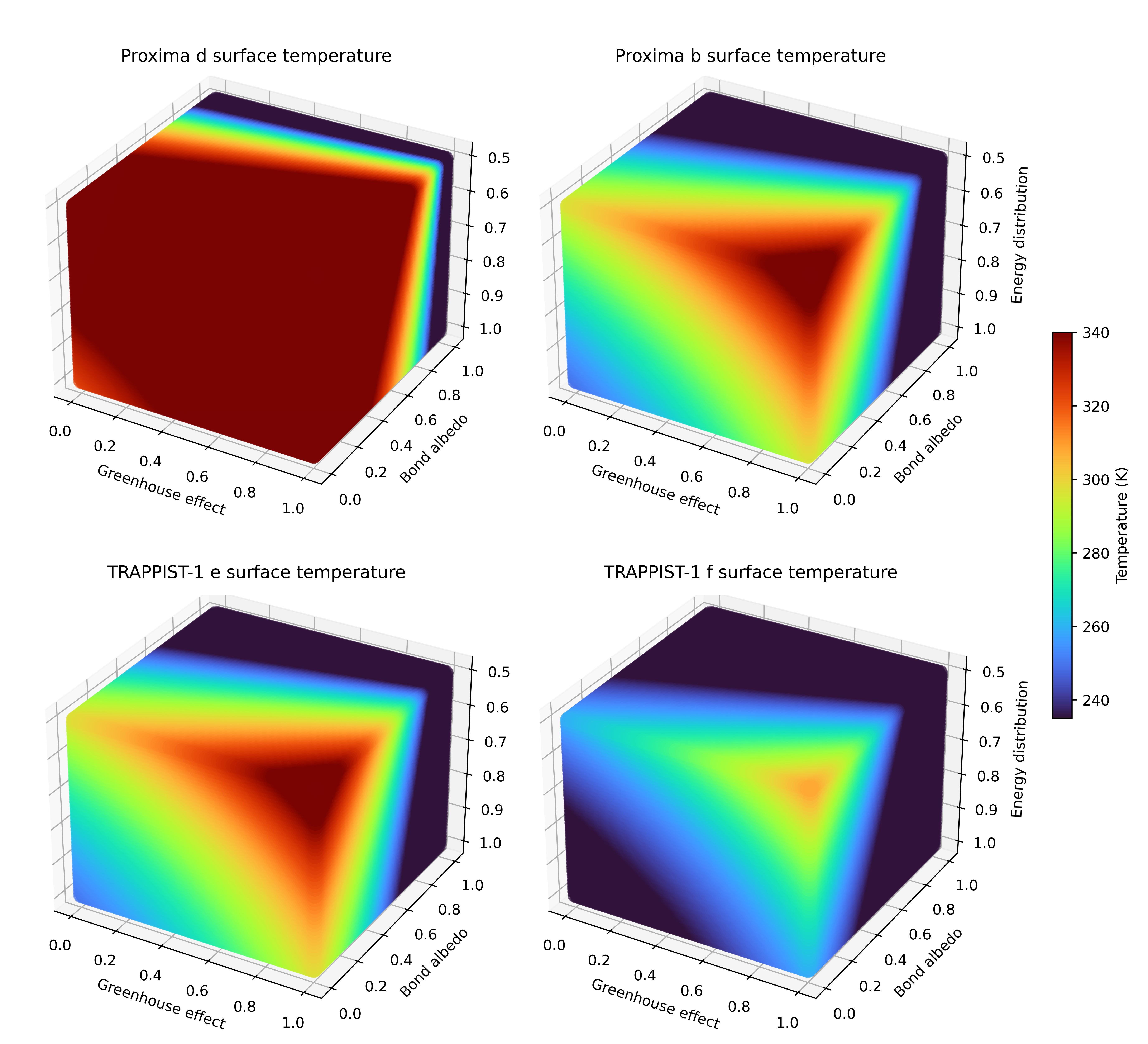}
	\caption{Calculated dayside surface temperature matrices for some of the studied planets. Temperatures are presented in Kelvin. $N$ = 1 was considered, since it is not expected that any of these planets have a very thick atmosphere. }
	\label{fig:fig4}
\end{figure}

    Our results suggest that Proxima b and TRAPPIST-1 e are the likeliest to have temperatures compatible with liquid water on their surface, and with the persistence of habitable environments. Proxima d and TRAPPIST-1 f are likely to have a too high and low dayside temperature, respectively. 
    
    Computing the model for all rocky planets of the Proxima and TRAPPIST-1 systems, assuming approximate values for factors $A_B$, $f$, $\varepsilon$ and $N$ based on current yields the results shown in Table 2. Earth-like values for all factors were used for Proxima b, TRAPPIST-1 e, and TRAPPIST-1 f, except $f=$ 0.75 (instead of 1) since these planets may be tidally locked, and, in this case, temperature dispersion is entirely dependent on the atmosphere and/or oceans. Proxima d was presumed to have a thin or non-existent atmosphere ($\varepsilon=$ 0) and thus minimal heat distribution ($f=$ 0.5) due to its very low mass and high incident stellar flux. TRAPPIST-1 d was assumed to have a dense Venus-like atmosphere and albedo. TRAPPIST-1 b \citep{greene2023} and c \citep{zieba2023} had their surface temperatures recently estimated by measurements from the James Webb Space Telescope, and those results were compared with our model. Cells are color-coded to represent the general expected habitability of a planet with such $T_d$. Cells in green (273 K $< T_d <$ 313 K) highlight ideal temperatures for supporting habitable environments on the surface. Cells in yellow (235 K $< T_d <$ 340 K) highlight temperatures within the limits of possible habitability, and cells in red show likely uninhabitable conditions \citep{godolt2016}. Table 2 also includes results for the Solar System’s planets and corresponding measurements found in the literature \citep{bauch2021, williamsplanetary2023}.

\begin{table}
	\centering
	\includegraphics{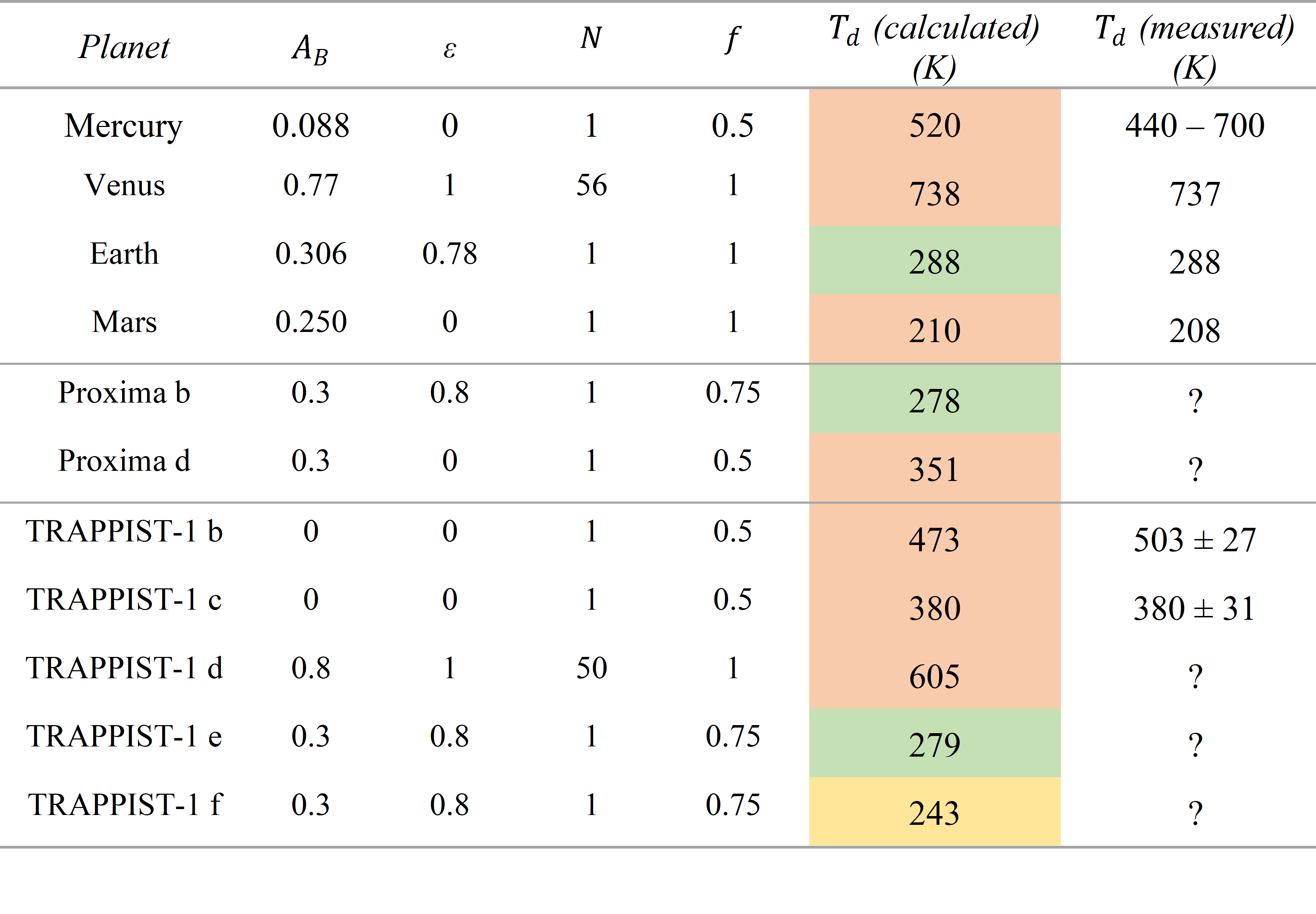}
	\caption{$f=$ 0.5 for Mercury due to its slow rotation and lack of atmosphere, creating a large temperature gradient between the day and night sides, due to the inefficient energy transfer.}
	\label{tab:tab2}
\end{table}

\subsection{Expected Radiation Environments}

    In Fig. 5 each planet is plotted according to its orbital distance against the calculated TOA X-ray (or UV) flux it receives due to an average flare. Table 3 shows the estimated doses of UV and X-rays during flares and superflares (energy $\geq$ 10$^{33}$ erg, \citealp{shibayama2013}).

\begin{figure}
	\centering
	\includegraphics[scale=0.8]{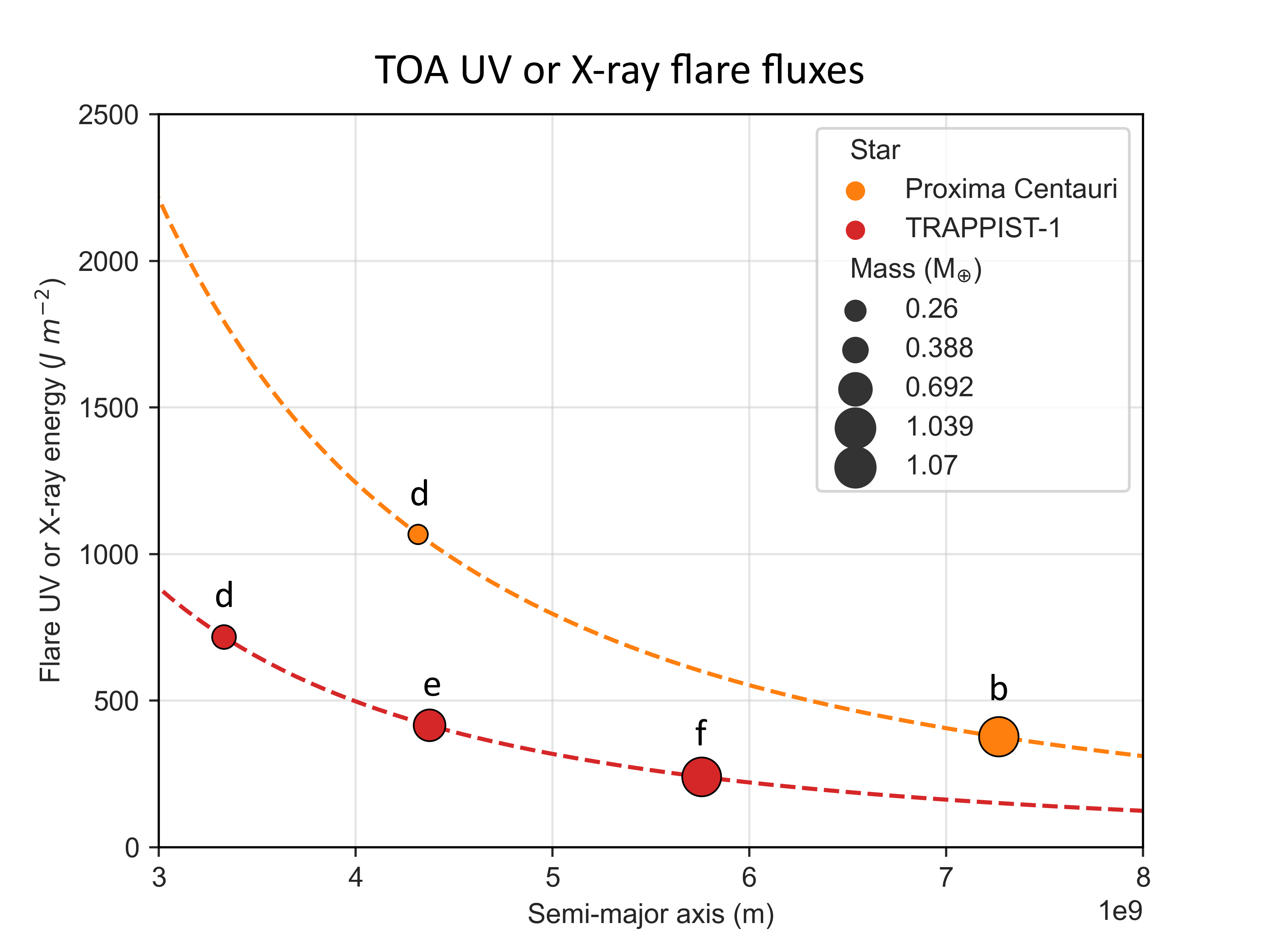}
	\caption{TOA UV or X-ray energy received by each studied planet during a flare with average output. The masses of the planets are also illustrated, as well as their distance to their respective star. Dashed lines represent the plotting of equation (1) with $E=$ 2.5$\times$10$^{30}$ erg (Proxima, orange) and $E=$ 1.0$\times$10$^{30}$ erg (TRAPPIST-1, red). }
	\label{fig:fig5}
\end{figure}

\begin{table}
	\centering
	\includegraphics[scale=0.6]{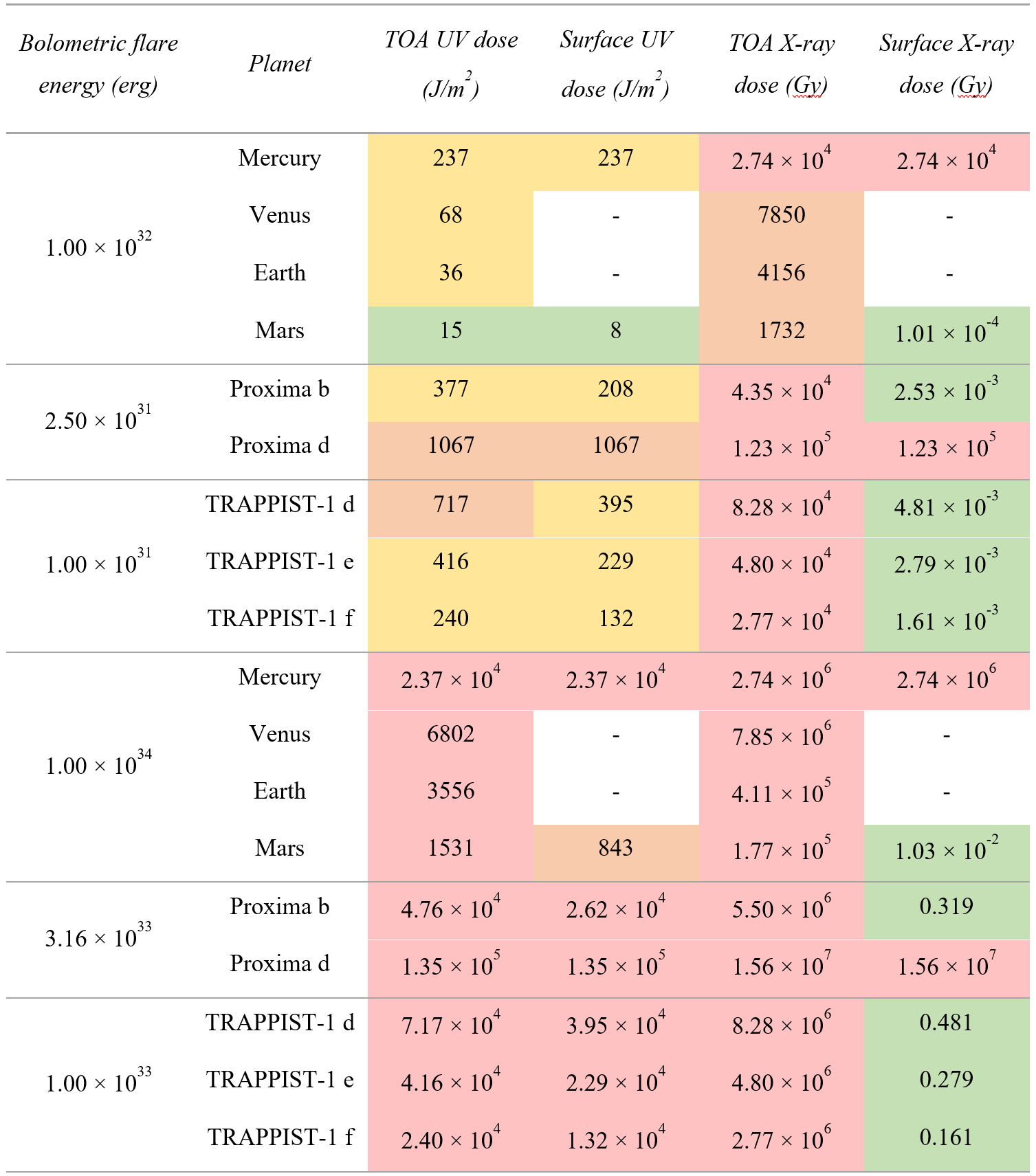}
	\caption{A Mars-like atmosphere with maximum transmittance was assumed to estimate the surface doses, except for Proxima d. Cells in green contain values below the LD$_{90}$ of \textit{E. coli}, \textit{A. niger} and \textit{D. radiodurans}; cells in yellow have values below the LD$_{90}$ of only \textit{A. niger} and \textit{D. radiodurans}; orange cells are only below one LD$_{90}$ (\textit{A. niger} or \textit{D. radiodurans}); and red cells are above all LD$_{90}$’s.}
	\label{tab:tab3}
\end{table}

    Bolometric flare energies are assumed as described in section 2.2.2. For superflares on Proxima and TRAPPIST-1, the bolometric energy was taken from the literature \citep{howard2018}, and the value for TRAPPIST-1 was generalized to be the minimum superflare energy since this star is less flare intensive \citep{yamashiki2019}. In M-dwarfs, the UV and X-ray flare output is similar, about 10\% of the bolometric energy output of the flare \citep{howard2018, welsh2007}. For the Sun, a bolometric flare energy of 10$^{32}$ erg was selected \citep{shibayama2013}, as well as a superflare energy of 10$^{34}$ erg \citep{shibata2013}. Superflares with this energy occur on the Sun around every 800 years \citep{shibata2013}. Although the Sun is a G-dwarf, the same 10\% fraction of the flare energy was assumed to be distributed in UV/X-rays. This is likely an overestimation of the UV and X-ray energy output during a solar flare \citep{reid2012, yamashiki2019} but provides a general comparison with the irradiance expected on the studied exoplanets. Using Mars’ atmosphere as a model, the surface doses were calculated using the modeled transmittance curves (Fig. 2 and 3), and these are shown in Table 3. Cells are colored according to the dose being above or below LD$_{90}$ (D$_{10}$) values. Note that these LD$_{90}$ values refer to more harmful UV-C irradiation, not the whole UV spectrum. Therefore, these represent a conservative estimate of the dose until a 90\% reduction in population is observed.
    
    As Table 3 emphasizes, a single flare event could sterilize the surface of mesophilic organisms, and mainly UV-hardy, spore-forming organisms such as \textit{A. niger} would survive. Even in the presence of a Mars-like atmosphere, surface doses remain too high for mesophilic survival. Furthermore, a superflare might eliminate most, if not all, organisms on the surface, due to the extremely high UV flux. However, atmospheric UV transmittance depends on many factors, including photon incidence angle, and the presence of dust or hazes. For instance, Proxima b is expected to have a high superflare UV dose, reaching a surface dose of 2.62$\times$10$^4$ J/m$^2$ with maximum transmittance, and 9127 J/m$^2$ corresponding to UV-C. But, in areas of intermediate transmittance (Fig. 3) the UV-C dose would be 1570 J/m$^2$, an elevated, but survivable value, at least for \textit{A. niger}, as shown in the next section. Moreover, for minimum transmittance, the superflare UV-C dose on Proxima b could be as low as 265 J/m$^2$, which is under the LD$_{90}$ of \textit{A. niger} and \textit{D. radiodurans}, as well as many other UV-tolerant organisms. On the planet with the weakest superflare irradiance, TRAPPIST-1 f, with maximum attenuation, UV-C doses only reach 134 J/m$^2$. Therefore, under certain conditions, even some superflares may not sterilize exoplanet surfaces of mesophilic microorganisms. For standard flares, although a minimal atmospheric attenuation scenario leads to total UV doses $>$100 J/m$^2$ in all exoplanets of this study, limiting the survival of mesophiles like \textit{E. coli}, maximal attenuation would reduce the doses of total UV to $\leq$53 J/m$^2$, and as low as 12 J/m$^2$, which would not create a significant hazard to most microorganisms. The subsurface environment could also shield cells from UV at even shallow depths since UV-C soil penetration depth is $\leq$0.11 mm \citep{ciani2005}.
    
    In the presence of an atmosphere, the same risk is not seen for X-rays, which, even for the strongest flares, generate low doses that do not pose a threat to microorganisms on the surface. Nonetheless, with no atmosphere present, the X-ray surface dose would be extremely elevated, potentially sterilizing the ground. In this scenario, some microorganisms could survive by living in the subsurface, where the dose very rapidly decreases at relatively shallow depths. For a superflare on Proxima b, the surface X-ray dose without an atmosphere is 5.50 $\times$ 10$^6$ Gy, but the dose below a 0.15 mm thick layer of soil (866 Gy) or water (9550 Gy) would be lower than the LD$_{90}$ extreme radiotolerant organisms like \textit{D. radiodurans} (Fig. 6). Moreover, at depths over 1 mm of soil, or 10 mm of water, most organisms – even mesophiles like \textit{E. coli} – could survive, meaning that subsurface and underwater environments would shield microorganisms from X-ray damage. 

\begin{figure}
	\centering
	\includegraphics[scale=0.8]{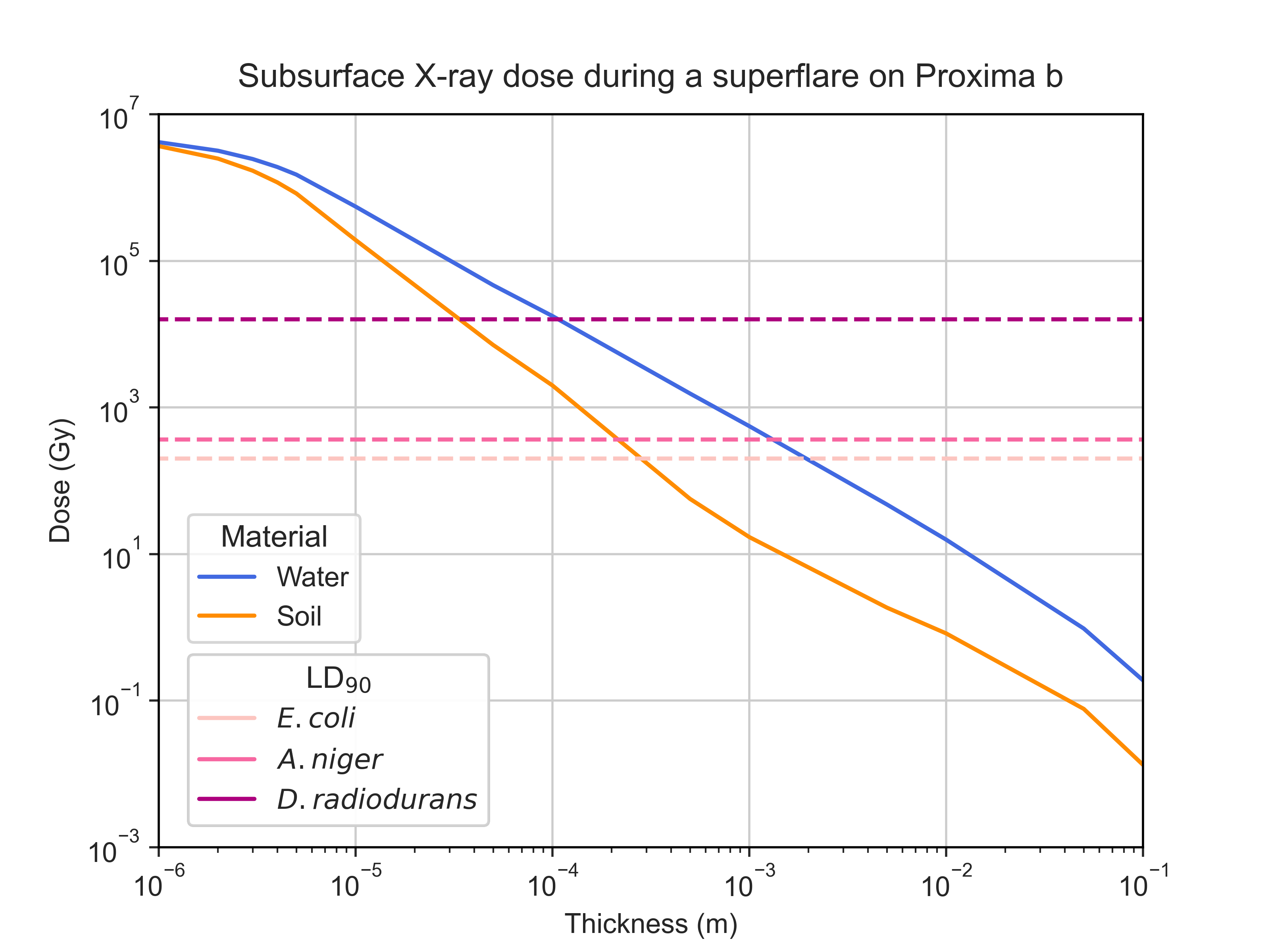}
	\caption{Estimated subsurface X-ray absorbed dose throughout a thin layer of soil (orange) or water (blue). Water has a lower capacity of attenuating these high-energy photons, leading to the need for a thicker layer to reduce the same dose when compared to soil. Dashed lines represent the LD$_{90}$ values for \textit{E. coli}, \textit{A. niger}, and \textit{D. radiodurans.} }
	\label{fig:fig6}
\end{figure}
    
    For an Earth-like atmosphere, the opacity to X-rays is essentially 100\%, and surface doses are negligible. There would also be increased UV protection, with UV-C being mostly eliminated, and UV-B drastically reduced \citep{segura2010}.

\subsection{Survival and Germination of \textit{A. niger} Spores to Flare UV and X-ray Irradiation and the Multifunctional Role of Melanin}

    Initial assays were performed to evaluate the outgrowth of wild-type and mutant \textit{A. niger} spores after X-ray irradiation (Fig. 7) considering a maximum dose of 1000 Gy, rounded up from the estimated Proxima b dose below 0.15 mm of soil (or 0.7 mm water) of 866 Gy mentioned in the previous section. Significant germination is observed with an irradiation of 100 Gy, but still a reduction when compared to the control, both for the wild-type (M-W, \textit{p} $<$ 0.00001) and the mutant (M-W, \textit{p} = 0.00008). To receive this dose, the spores would need shielding of 0.4 mm of soil or 3 mm of water. At 500 Gy (0.2 mm soil, 1 mm water) some germination of the wild-type is observed, significantly more than the mutant (M-W, \textit{p} = 0.005). At the largest dose of 1000 Gy little germination is seen in any strain. 

\begin{figure}
	\centering
	\includegraphics[scale=0.6]{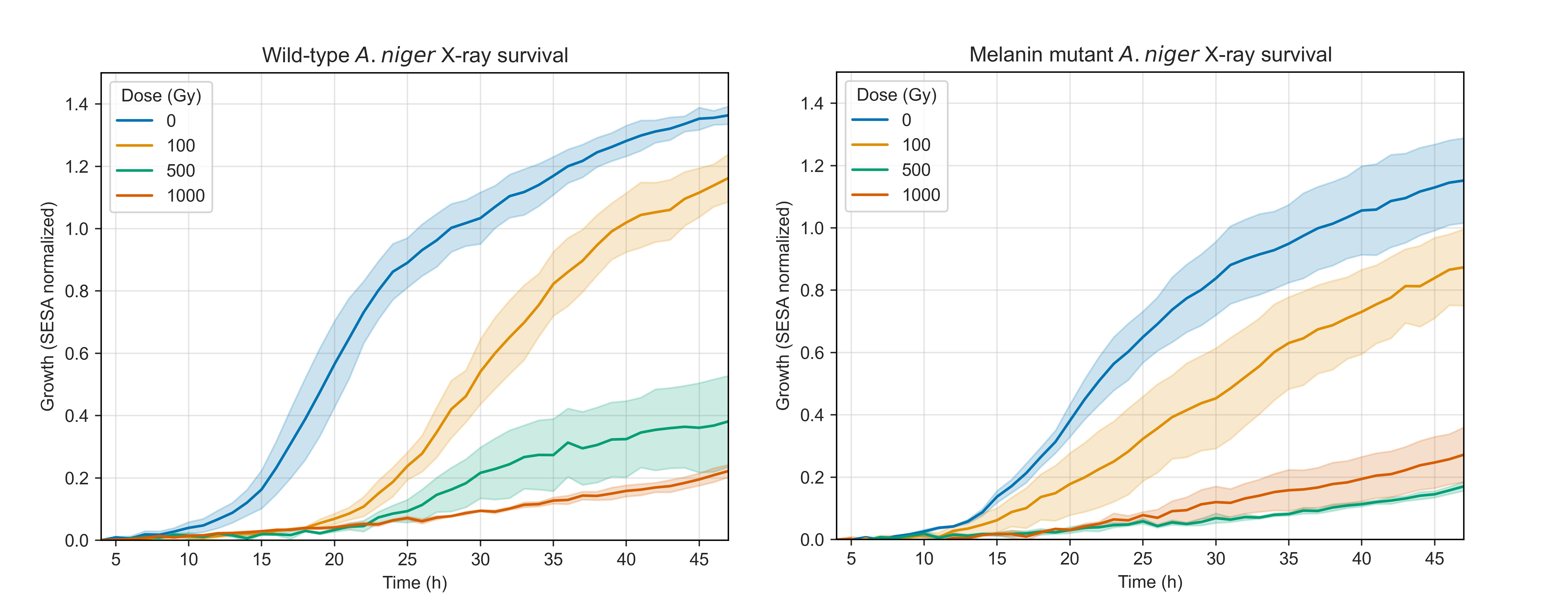}
	\caption{Spore germination of \textit{A. niger} wild-type (left) and mutant (right) strains after X-ray irradiation, measured through the SESA fungi algorithm of the oCelloScope\texttrademark. Uncertainty bands represent the standard error over the replicates (n = 3). The normalization was done at t = 4h as detailed in the Methods section. }
	\label{fig:fig7}
\end{figure}
    
    To test the protective efficiency of a solid melanin layer vs. the melanin-rich solution produced by the growth of OS4.3, wild-type, and mutant spores were irradiated with 1000 Gy of X-rays, suspended in either a standard saline solution or in the filtered melanin-rich solution. Three samples suspended in 0.9\% NaCl were under a thin solid melanin film. As seen in Fig. 8, for both the wild-type and the mutant (K-W, \textit{p} $<$ 0.00001) there is greater germination for irradiated spores suspended in the melanin-rich solution, whereas no shielding effect is seen for the solid melanin film, both on the wild-type and mutant. Better germination of the mutant controls compared to the wild-type controls is likely due to experimental error and variability as opposed to physiological effect, particularly since Fig. 7 shows the opposite tendency. In any case, the melanin-rich non-irradiated controls grew better than their saline solution counterparts in the wild-type (M-W, \textit{p} $<$ 0.00001), but not significantly so for the mutant (M-W, \textit{p} = 0.52), although a tendency for faster germination in melanin can be seen. Fig. 9 shows images of samples in 0.9\% NaCl and in the melanin solution at 5 and 48 hours after inoculation in MM. 

\begin{figure}
	\centering
	\includegraphics[scale=0.6]{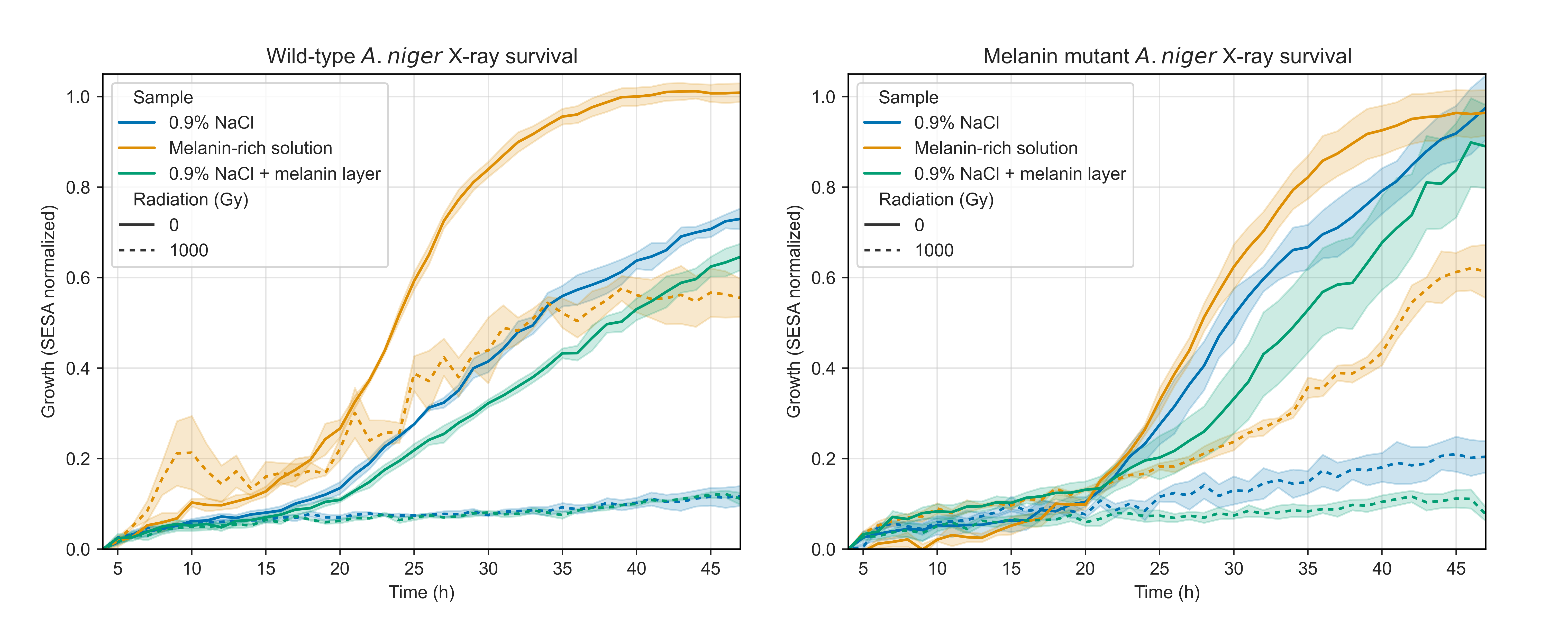}
	\caption{Spore germination of \textit{A. niger} wild-type (left) and mutant (right) strains in two different solutions. The protective effect of a solid melanin layer was also tested on 3 samples in saline solution. Data presented shows irradiated (1000 Gy X-rays, dashed lines) and control samples (non-irradiated, solid lines). }
	\label{fig:fig8}
\end{figure}

\begin{figure}
	\centering
	\includegraphics[scale=0.8]{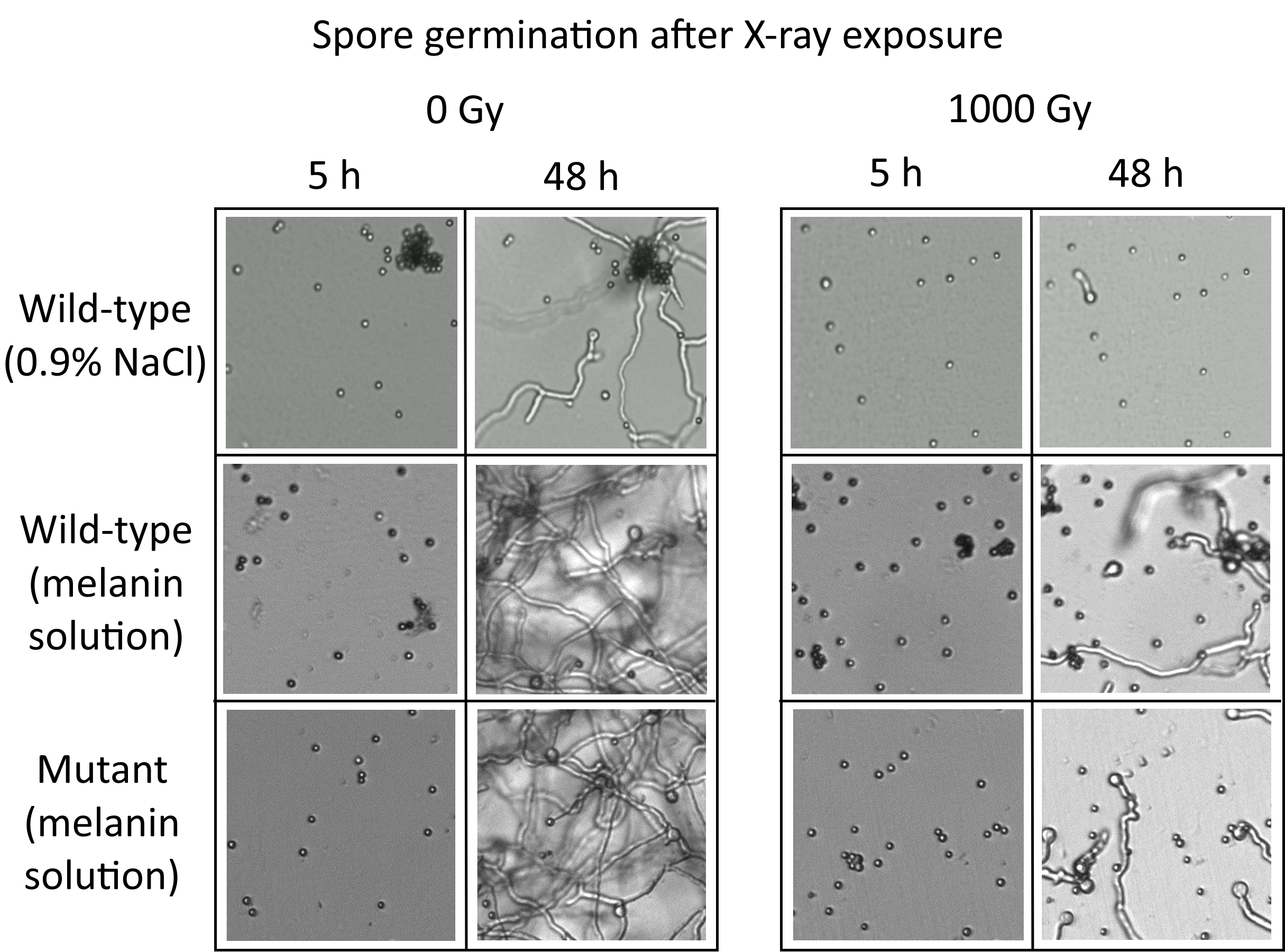}
	\caption{Images from the germination of \textit{A. niger} spores, in a saline solution (0.9\% NaCl), or a filtered supernatant containing solubilized pyomelanin. Control and irradiated (1000 Gy of X-rays) samples of the wild-type and mutant strains are shown.  }
	\label{fig:fig9}
\end{figure}

    After these tests, further experiments were performed to better characterize the beneficial effects of suspension in a melanin-rich solution for both UV-C and X-ray irradiation shielding and spore recovery after irradiation. Survival assays showed that melanin-deficient \textit{A. niger} MA93.1 spores survive 5000 J/m$^2$ of UV-C (254 nm) if suspended in a control supernatant or in a melanin-rich supernatant (Fig. 10). Notably, survival is higher when melanin is present (t-test, \textit{p} = 0.006), since an average of 44\% of spores in melanin survived the highest dose, whereas 2.7\% survived in the control supernatant. Survival fraction change is negligible between 1000 and 5000 J/m$^2$, which is not the case for the control supernatant (with fungal extracellular compounds, but without melanin). In contrast, spores suspended in saline solution were significantly more susceptible to UV-C, even at 1000 J/m$^2$ (t-test, \textit{p} = 0.01), and less than 1\% survived until 2500 J/m$^2$. No survival was observed at the highest tested dose.

\begin{figure}
	\centering
	\includegraphics[scale=0.6]{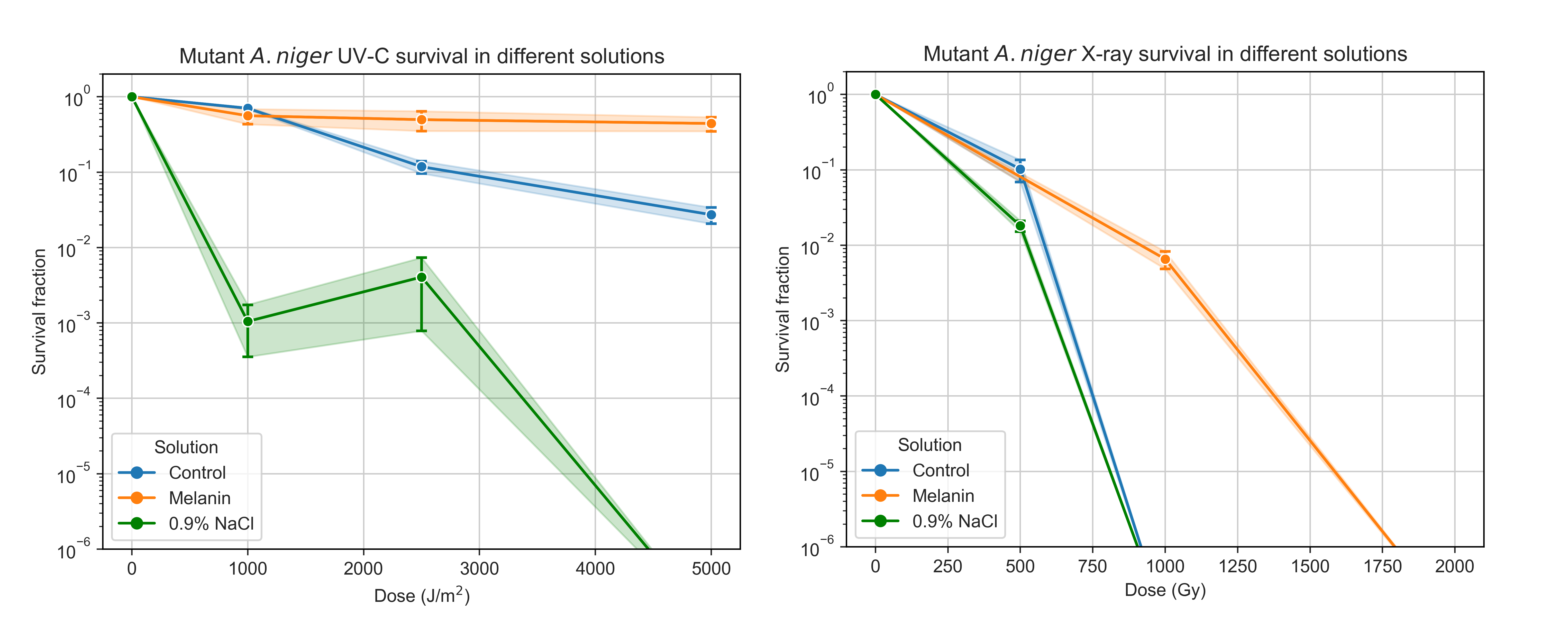}
	\caption{Survival fractions of \textit{A. niger} MA93.1 spores when exposed to UV-C (left) and X-ray radiation (right) in three distinct solutions, a 0.9\% NaCl solution, a melanin-free supernatant (“Melanin”), and an identical but melanin-rich supernatant (“Control”).  }
	\label{fig:fig10}
\end{figure}

    Spores of the melanin-deficient strain irradiated with X-rays showed a similar dose-response pattern, with only the spores suspended in the melanin solution surviving a dose of 1000 Gy, although the inactivation fraction was $>$99\%. Control samples show improved survival at 500 Gy when compared to 0.9\% NaCl samples (t-test, \textit{p} = 0.03), but in both cases no spores survived higher doses ($\geq$1000 Gy). No samples survived a 2000 Gy irradiation. 
    
    Fig. 11 suggests that spores in melanin solutions germinated faster and more efficiently than those in saline solution or a control supernatant, both for irradiated and non-irradiated samples, indicating that extracellular melanin could be a spore germination trigger.
    
\begin{figure}
	\centering
	\includegraphics[scale=0.6]{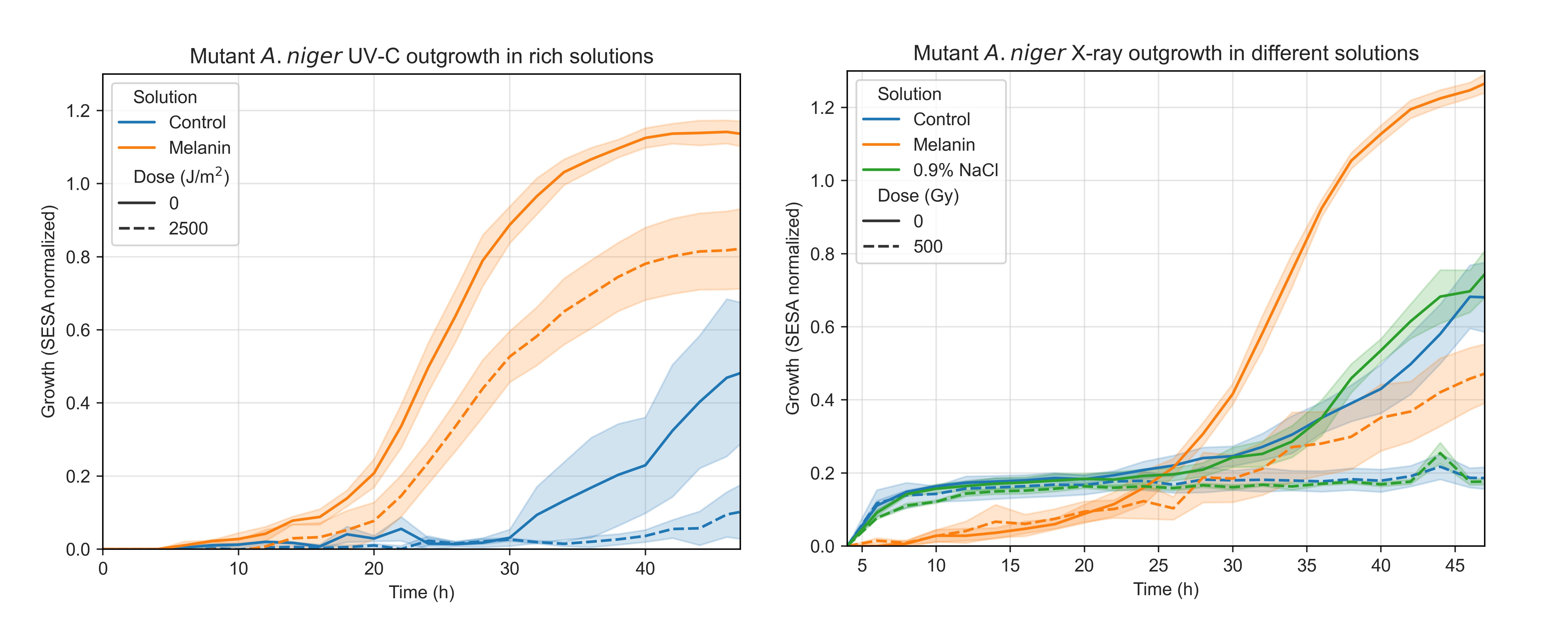}
	\caption{Left – Measured germination of \textit{A. niger} melanin mutant (MA93.1) in two different supernatant solutions, one melanin-rich (“Melanin”) and another free of melanin (“Control”). Data presented shows irradiated (2500 J/m$^2$ UV-C, dashed lines) and control samples (non-irradiated, solid lines). Right – Measured germination of A.niger melanin mutant (MA93.1) three different solutions. Data presented shows irradiated (500 Gy X-rays, dashed lines) and control samples (non-irradiated, solid lines). }
	\label{fig:fig11}
\end{figure}

\section{Discussion}
\subsection{Models of (Exo)planetary Surface Temperature and Radiation Environment }

    We developed a robust but streamlined 0D model to estimate a planet’s dayside surface temperature. Complex models, like general circulation models, although potentially more accurate, require a greater amount and precision of information which is largely unknown for exoplanets. Equation (1) is not intended to be an alternative to 1D-3D models, but rather an efficient tool for when not enough information is present and rapid, general estimations under various conditions are intended. This allows astrobiologists to construct an overview of different possibilities to then plan experiments on simulated exoplanet environments. 

    Furthermore, we developed versatile models to estimate the flare TOA, surface and subsurface doses of UV and X-rays on exoplanets, and their effect on model microorganisms, by using the Martian atmosphere as a model. The atmospheric attenuation model for X-rays does not account for various physical processes, such as photon interactions, secondary production, and charge-particle equilibrium, and only vertical column density is factored in, neglecting horizontal attenuation. When modeling UV attenuation, the used equations were based on the generalization of the Martian atmospheric transmittance under three sets of conditions, and thus do not adequately characterize all possible combinations of atmospheric traits. Nonetheless, the developed equations are useful to inform microbiological experiments for exoplanet flare irradiation studies. 

\subsection{Are Proxima b and TRAPPIST-1 e Good Candidates for (Sub)Surface Habitability?}

    Estimated dayside temperatures for Proxima b and d, and TRAPPIST-1 d, e and f are in line with the values estimated from more complex models \citep{boutle2017, lincowski2018, sergeev2020, turbet2016, wolf2017, wunderlich2020}. According to our results, Proxima b and TRAPPIST-1 e are good candidates to have potentially temperature environments, enhancing their habitability. 

    Planetary radiation doses depend on stellar flare energy and planetary traits (orbit and atmosphere). If unattenuated, X-rays from flares would most likely sterilize the surface of all studied exoplanets. However, microorganisms suited to survive under the surface would be unaffected by most exogenous radiation sources under a few millimeters of soil or water. 

\subsection{Melanin-rich Solutions Increase the Survival and Germination of \textit{A. niger} Spores}

    The experiments performed in this study corroborate the multifunctional purpose of melanin since \textit{A. niger} MA93.1 spores germinated faster and more efficiently in a melanin-rich extract when compared to the two control solutions. Non-irradiated spores in both control solutions showed similar outgrowth capability (Fig. 8). Both untreated and irradiated spores in the melanin-rich solution showed greater germination capacity than the controls, including the formation of more complex hyphal networks. Outgrowth capacity is connected not only to the survival rate but also to the rate of DNA repair. Even if the survival is similar, faster germination indicates less DNA damage, since Aspergillus spores that detect altered DNA inhibit the germination process until repairing is complete \citep{harris1998, ye1997}. Thus, the DNA repair process was quicker, and germination started sooner. 

    For irradiated spores, a clear increase in survival is seen in samples containing melanin. For UV-C, nearly half of the spores survived a 5000 J/m$^2$ dose, comparable to a superflare on TRAPPIST-1 f with minimum atmospheric attenuation (4602 J/m$^2$ of UV-C), indicating that a significant fraction of spores in these conditions could survive superflares on the exoplanets more likely to be habitable – Proxima b (9127 J/m$^2$) and TRAPPIST-1 e (7976 J/m$^2$) – even with minimal atmospheric shielding. Finally, for X-rays, spores in the control solutions survived only up to 500 Gy, while samples in the melanin-rich solution showed survival up to 1000 Gy, as well as increased germination capacity in all scenarios.
    
    The functional diversity of microbial melanins is notable, as they have been associated with a variety of roles \citep{cordero2017}. However, no prior research had found a correlation between solubilized melanin and spore development efficiency. To explain our results, we suggest that the effect of the presence of melanin is two-fold. Firstly, for irradiated samples, it serves a shielding purpose, protecting the spores from direct damage from incoming radiation. Secondly, the presence of melanin in the medium may be advantageous in cellular processes that could promote the upkeep of a high spore viability and growth efficiency, for example through reactive oxygen species scavenging, or by triggering quorum sensing pathways related to germination. Although the production of melanin is regulated by quorum sensing in fungal species \citep{albuquerque2013, homer2016}, no studies have identified melanin as participating as an inducer. Finally, melanin could have functions that directly impact spore germination, hyphal morphogenesis, and growth rate, as demonstrated in \textit{A. niger} \citep{cortesao2022} and other fungal species (e.g. \citealp{yu2015}). In any case, follow-up experiments are required to better explain the results obtained in this work, including testing the wild-type strain, and performing detailed biochemical characterization of the melanin-rich solution.

\section{Conclusion}
    
    Overall, the work developed during this study highlights the advantages of applying an interdisciplinary approach to astrobiology and exoplanet science. The lack of detailed information about exoplanet surfaces has led astrobiologists to resort to useful, but often subpar measures of habitability, such as the equilibrium temperature and the Goldilocks zone. As we showed during this work, a combined approach leveraging astrophysical modeling and observational extrapolation of exoplanet conditions (e.g. temperature, radiation environment) with microbiological experiments may allow astrobiologists to construct hypothetical, but realistic, model environments on which to test microbial survival and growth. Furthermore, results from this work showed how \textit{A. niger}, like other extremotolerant and extremophilic organisms, would be able to survive harsh radiation conditions on the surface of some M-dwarf exoplanets. Additionally, melanin-rich solutions were shown to be highly beneficial to the survival and germination of \textit{A. niger} spores, particularly when treated with high doses of UV and X-ray radiation. These results offer an insight into how lifeforms may endure harmful events and conditions prevalent on exoplanets, and how melanin may have had a role in the origin and evolution of life on Earth, and perhaps on other worlds.

\vspace{0.6cm}
\textbf{Acknowledgements}

    We thank Dr. Marco Moracci (University of Naples “Federico II”) for his guidance and assistance during the project. We are also grateful to Héctor Palomeque for his help with modeling and statistical analysis.

\vspace{0.6cm}
\textbf{Author Contributions}

    AM: conceptualization, data curation, formal analysis, investigation, methodology, project administration, visualization, writing – original draft, writing – review and editing. SK: methodology, resources, validation, writing – review and editing. DM: methodology, validation, writing – review and editing. NS: supervision, writing – review and editing. MC: conceptualization, methodology, project administration, resources, supervision, writing – review and editing.

\vspace{0.6cm}
\textbf{Author Disclosure Statement}

    The authors declare no competing interests.

\vspace{0.6cm}
\textbf{Funding Statement}

    The study was supported by DLR internal funds.

\vspace{0.6cm}

\bibliography{main}

\vspace{2cm}
\textbf{Supplementary Material}

\renewcommand{\theequation}{S\arabic{equation}}
\setcounter{equation}{0}
\renewcommand{\thefigure}{S\arabic{figure}}
\setcounter{figure}{0}
\renewcommand{\thetable}{S\arabic{table}}
\setcounter{table}{0}

    The light flux that reaches an object is defined as $F$, calculated using the star’s temperature ($T_{\star}$) and radius ($R_{\star}$), as well as the planet’s semi-major axis (i.e. its average orbital distance to the star, $a$), as shown in equation (S1):

\begin{equation}
    F=\frac{\sigma R_{\star}^2T_{\star}^4}{a^2}
\end{equation}

    The planetary equilibrium temperature ($T_{eq}$) is the temperature that a planet would be at if it perfectly absorbed all incoming radiation from its star and re-radiated it uniformly across its entire surface, assuming no internal heat sources, no atmospheric effects, and complete redistribution of energy across the planet. Based on the Stefan-Boltzmann law, a planet’s $T_{eq}$ can be calculated with equation (S2) (e.g., \citealp{seager2011}):

\begin{equation}
	T_{eq}=\sqrt[4]{\frac {F(1-A_B)}{4\sigma}}
\end{equation}

    Here, $F$ is the stellar flux received by the planet, $A_B$ is its Bond albedo. The factor 4 comes from the fact that the total solar energy incident on a planet directly illuminates its cross-sectional area, a circle with area $\pi r^2$. However, the assumption done is that this energy disperses across its entire spherical surface, $4\pi r^2$.
    
    Equation (S2) is useful to determine the equilibrium temperature of planets with orbits similar to those of the Solar System’s planets. But some stars, including M-dwarfs, have planets orbiting in much closer proximity, leading to tidal locking. In addition, some exoplanets may exhibit more varied orbital resonance behaviors. To account for these factors, a modification can be made to equation (S2) to weight in the planet’s heat distribution (e.g. \citealp{seager2011}):
    
\begin{equation}
	T_{eq}=\sqrt[4]{\frac {F(1-A_B)}{4f\sigma}}
\end{equation}

    The variable $f$ represents the fraction of the planet’s surface area over which the heat is distributed. Thus, none of the heat is spread to $(1-f)$ of the surface area. In tidally locked or slowly rotating planets, $f=$ 0.5, with a non-existent heat transfer from the illuminated side to the dark side, and therefore, in this case, $T_{eq}$ is estimating the temperature of the lit side, while the dark side is extremely cold (i.e. $T_{eq}=T_d$). For rapidly rotating planets, $f=$ 1, and, in that case, equation (S3) is equivalent to equation (S2), as expected.
    
    The equilibrium temperature is insufficient to properly evaluate a planet’s habitability. For instance, Earth’s $T_{eq}$ is around 255 K (-18 ºC), which is lower than observed. This temperature would be too cold for the survival of most organisms, as it is well below the freezing point of water. Nonetheless, Earth’s surface temperature is around 33 K higher \citep{williamsplanetary2023}, and this discrepancy can be explained by the atmospheric greenhouse effect. Accurately modelling how greenhouse gases affect a planet’s climate (and, therefore, temperature) is complex, but some assumptions and simplifications can be made while still obtaining fairly accurate predictions. A commonly used approximation is a one-layer atmosphere model, where a planet’s atmosphere is considered only as a single layer of gas with albedo A that can either absorb or emit the stellar irradiance ($\frac{F}{4}$) it receives. Assuming all wavelengths of infrared radiation are absorbed/emitted equally well, the atmosphere can be treated as a blackbody \citep{kump2014}. Fig. S1 illustrates the concept of this model, as well as the general avenues for heat loss or retention.

\begin{figure}
	\centering
	\includegraphics[scale=0.4]{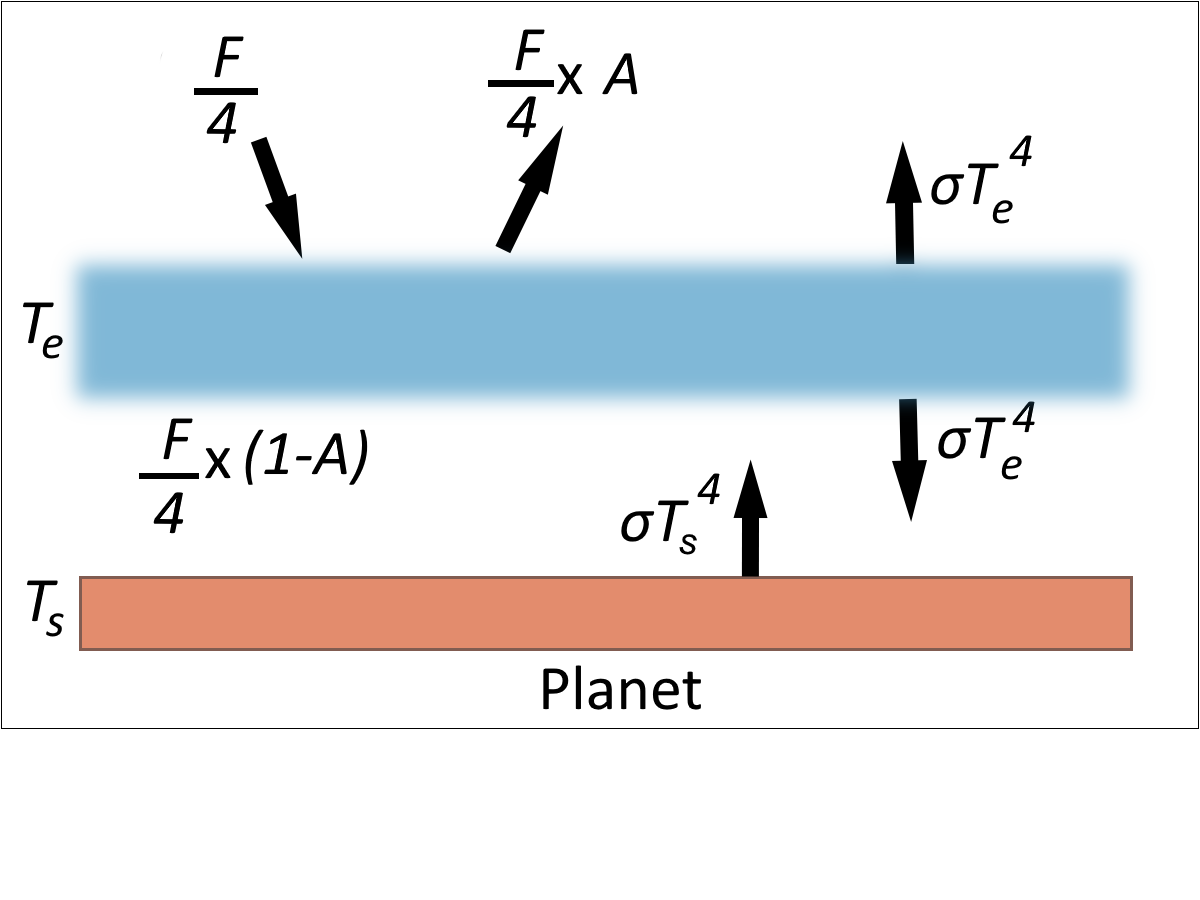}
	\caption{Schematic overview of a one-layer atmosphere and the energy flux within. $\sigma$ is the Stefan-Boltzmann constant, and $\sigma T^4$ values represent the emission of infra-red (IR) radiation (i.e., heat). $T_e$ is the atmospheric temperature and $T_s$ the surface temperature. }
	\label{fig:figS1}
\end{figure}

    Using this model, it is possible to calculate a more accurate value for a planet’s surface temperature, depending on its atmosphere, as shown in equation (S4) \citep{kump2014}:

\begin{equation}
        T_s=\sqrt[4]{\frac {F(1-A_B)}{4\sigma(1-\frac{\varepsilon}{2})}}
\end{equation}

    Here, $\varepsilon$ is a value between 0 and 1 that represents the fraction of IR radiation emitted by the surface ($\sigma T_e^4$) that is absorbed by the atmosphere – the atmospheric greenhouse effect. 
    When applying equation (S4) to the Solar System’s rocky planets, this model is able to accurately predict surface temperature of Mercury, Earth and Mars, given the corresponding $\varepsilon$ values, as demonstrated in Table S1. However, Venus’ calculated temperature is much lower than the measured surface temperature. This is due to the single-layered nature of the model, which is incapable of correctly predicting exceedingly dense atmospheres with a very strong greenhouse effect, as is the case for Venus.

\renewcommand{\arraystretch}{1.5}
\begin{table}[]
    \centering
    \begin{tabular}{|c|c|c|c|}
         \hline
         Planet & $T_s$ (literature) (K) & $\varepsilon$ & $T_s$ (calculated) (K) \\
         \hline
         \hline
         Mercury	& 440 & 0 & 438 \\
         \hline
         Venus & 737 & 1 & 273 \\
         \hline         
         Earth & 288 & 0.78 & 288 \\
         \hline         
         Mars & 208 & 0 & 210 \\
         \hline
    \end{tabular}
    \caption{Observed and calculated temperatures for the rocky planets in the Solar System using equation S4. $\varepsilon=$ 0 was assumed for Mercury and Mars, due to their thin atmospheres, which provide little to no greenhouse effect. Earth is known to have $\varepsilon$ between 0.77 \citep{jaboc1999} and 0.79 \citep{liu2020}, while Venus’ thick atmosphere creates a very strong greenhouse with $\varepsilon \approx$ 1 \citep{liu2020}. Literature $T_s$ values were taken from \citet{williamsplanetary2023}}. 

\end{table}

    To correctly predict the surface temperatures of Venus-like planets with extremely dense atmospheres, a $N$ number of layers can be considered, instead of only one, which increases the greenhouse capabilities of the modelled atmosphere, similarly to what was shown in \citet{liu2020}:

\begin{equation}
	T_d=\sqrt[4]{\frac {N \cdot F(1-A_B)}{4\sigma(1-\frac{\varepsilon}{2})}}
\end{equation}

    Using equation (S5), with $N=$ 56, the calculated surface temperature for Venus is 738 K, equivalent to the measured $T_s$ \citep{williamsplanetary2023}, while $N=$ 1 yields the correct results for the rest of the rocky planets shown. 
    
    Nevertheless, as previously discussed for the $T_{eq}$, tidally locked and orbitally resonant planets must also be considered, since these are likely to be present in many M-dwarf systems. To account for this, a factor $f$ can be used, as described for equation (S3). With it, a final model for predicting surface temperatures of rocky (exo)planets on the dayside emerges as:

\begin{equation}
	T_d=\sqrt[4]{\frac {N \cdot F(1-A_B)}{4f\sigma(1-\frac{\varepsilon}{2})}}=T_{eq} \cdot \sqrt[4]{\frac {N}{f(1-\frac{\varepsilon}{2})}}
\end{equation}

    Figure S2 shows this equation applied to TRAPPIST-1 d under an assumed "standard" atmosphere ($N=$ 1) and a Venus-like atmosphere ($N=$ 50):

\begin{figure}
	\centering
	\includegraphics[scale=0.8]{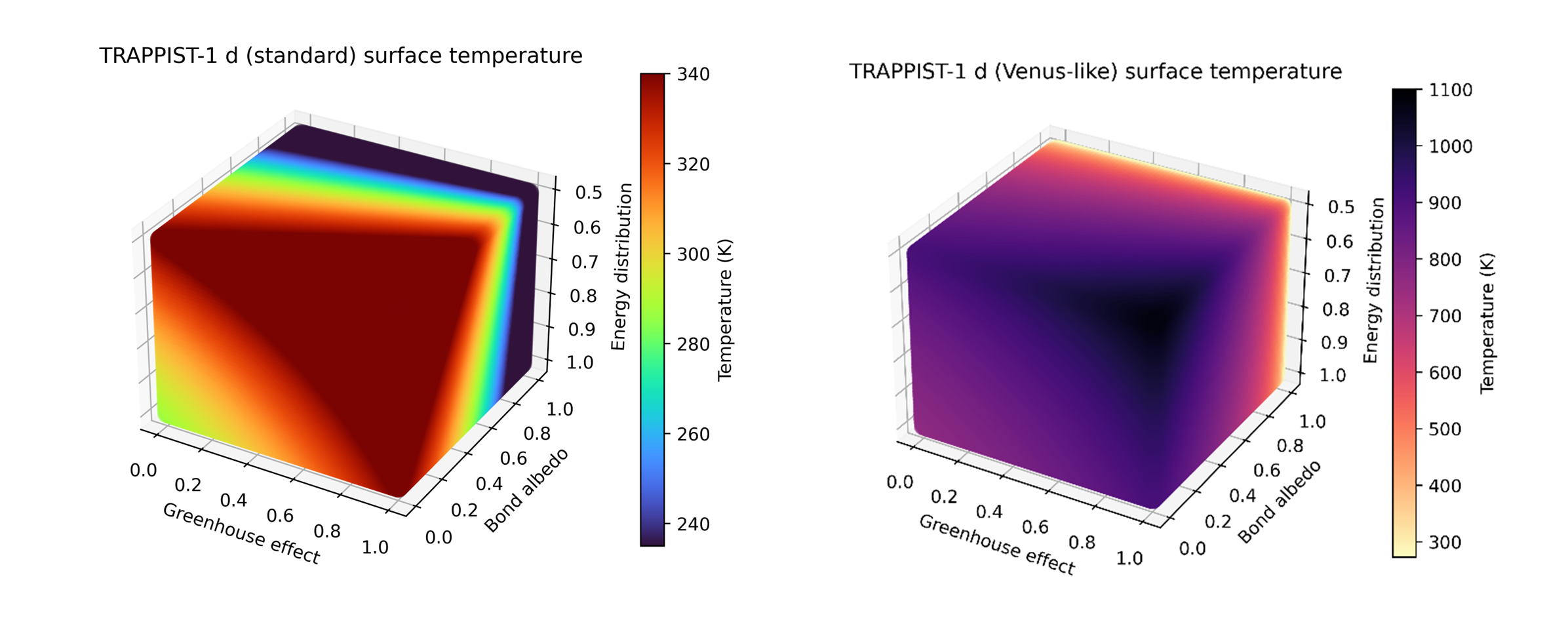}
	\caption{Estimated surface temperature matrix of TRAPPIST-1 d for a standard atmosphere (left) and a thick Venus-like atmosphere (right). }
	\label{fig:figS2}
\end{figure}

\end{document}